# Exciton Coherence in $CsPbBr_3$ Nanocrystals is Bounded by Phonon-Mediated Bright-Triplet Relaxation

**Authors:** Tara Šverko[1], Annette J. Jones[1], Chantalle J. Krajewska[1], Peter C. Sercel[2,3], Alexander L. Efros[4], Alexander E. Kaplan[1], Niamh L. Brown[1], Stefano Toso[4,5], Moungi G. Bawendi[1]*

**Affiliations:**

[1]Department of Chemistry, Massachusetts Institute of Technology; Cambridge, MA, 02139, USA.

[2]Center for Hybrid Organic Inorganic Semiconductors for Energy; Golden, CO, 80401, USA.

[3]National Laboratory of the Rockies, Golden, CO, 80401, USA

[3]Center for Computational Materials Science, U.S. Naval Research Laboratory; Washington, DC, 20375, USA

[4]Department of Chemical Engineering, Massachusetts Institute of Technology; Cambridge, MA, 02139, USA.

[5]Department of Chemistry, Lund University, Box 124, Lund, Sweden, SE-221 00

*Corresponding author. Email: mgb@mit.edu.

**Abstract**: Scalable sources of indistinguishable single photons or entangled photon pairs are fundamental to many quantum photonic technologies. Colloidal lead halide perovskite nanocrystals are promising such sources, but their exciton coherent properties are not fully understood. We show that in single $CsPbBr_3$ nanocrystals at 4 K, acoustic phonon-mediated exciton fine structure relaxation (EFSR) drives leakage of population between the bright exciton triplet states, competing with the radiative lifetime. The leakage pathway reaches $0.64 \pm 0.04$ of the radiative rate, bounding state coherence to 0.81 times the transform limit at 4 K. This pathway is intrinsic to the nanocrystal, not its environment, making it a tractable target for materials and device design. We identify fine structure and acoustic phonon engineering as effective intrinsic routes to higher coherence in perovskite quantum light sources.

**Main Text:**

Photon-based quantum information technologies, including quantum computing and quantum cryptography, rely on the ability to generate pure quantum states of light on demand.(*1, 2*) These sources must produce photons that can be manipulated in a controlled manner, such as highly indistinguishable single photons, or directly generate correlated or entangled photon pairs.(*3*) To date, semiconductor quantum dots (QDs) represent some of the best-performing solid-state platforms for quantum light generation due to their versatility as “artificial atoms.”(*4*) Unlike isolated atoms, QDs offer more synthetic degrees of freedom, such as wavefunction engineering through quantum confinement and easier integration into optical cavities.(*3, 5*) QDs synthesized by molecular beam epitaxy have demonstrated high-performance single-photon emission across a broad spectral range,(*6-8*) as well as the generation of entangled photon pairs via the biexciton-exciton cascade.(*3, 9, 10*) Colloidal QDs, a more versatile class, are synthesized through bottom-up growth from molecular precursors; their synthesis, purification, and separation bridge molecular and colloidal chemistry.(*11*) Advances in synthesis have enabled remarkable variety in size and morphology across a large library of compositions, including II-VI, III-V, ternary, and metal halide perovskites. Despite the nucleation and growth of billions of particles, these methods achieve high particle monodispersity and emission from the ultraviolet to the infrared.(*11*) Surface ligands further allow processing and purification in a wide range of media and tune the electronic properties at the semiconductor surface.(*12*)

Compared to isolated atoms, QDs couple more strongly to their solid-state and ligand environment, which can lead to variability of emitted photons.(*5, 11*) The excited state, or exciton, is described by a Bloch wavefunction which carries the underlying symmetry of the crystal lattice.(*13*) Electron-hole exchange interactions further split the exciton into fine structure states that can undergo population transfer. This process, known as exciton fine structure relaxation (EFSR), is a nonradiative transition that can be mediated by phonons or spin noise.(*14*) The exciton may undergo pure dephasing through fluctuating external charges, interactions with surface ligands, or environmental noise. Both mechanisms contribute to decoherence of the exciton: EFSR drives transitions between populations, while pure dephasing fluctuations scramble the phase within a population.(*3, 9, 15, 16*) Reaching the high photon indistinguishability required for quantum photonic applications means closing the gap between the coherence time and radiative lifetime, that is, reaching the transform limit.(*17, 18*)

Among colloidal QDs, only lead halide perovskite nanocrystals (LHP NCs) have shown performance approaching that of established quantum-photonic light sources.(*4, 11*) Weakly confined $CsPbBr_3$ nanocrystals have exhibited Hong-Ou-Mandel experiments visibilities up to $0.56 \pm 0.12$ even without Purcell-type radiative enhancement.(*19*) However, this measurement does not distinguish between pure dephasing and population relaxation. We directly measure the contribution of population relaxation to decoherence in LHP NCs by leveraging the biexciton-exciton cascade.

Monitoring exciton relaxation through polarization-resolved second-order correlations, we map the timescales and pathways of EFSR directly. We show that phonon-assisted EFSR is a significant intrinsic leakage channel that limits coherence, with the zero-temperature EFSR rate constant reaching $0.64 \pm 0.04$ times the radiative rate. In the best nanocrystal measured, this single pathway limits coherence to 0.81 times the transform limit at 4 K. The pathway is intrinsic to the material rather than coupling to the environment and thus can be mitigated effectively through rational material design or device-level engineering.

**Nanoparticle synthesis**

We synthesized weakly confined $CsPbBr_3$ nanocrystals (NCs) using an established hot-injection method.(*19, 20*) The room-temperature ensemble absorption spectrum shows an excitonic feature at 2.412 eV, characteristic of cuboidal $CsPbBr_3$ NCs with edge length 17 nm (Fig. 1A). The ensemble emission spectrum displays a single peak at 2.939 eV with a narrow linewidth of 75 meV (Fig. 1A), approaching the room-temperature single-particle linewidth reported for this material.(*21, 22*) HAADF-STEM images confirm the edge length and reveal that the as-synthesized NCs are quasi-cubic (Fig. 1B, fig. S1). XRD measurements confirm the particle size and cubic shape (fig. S2-3), while the underlying lattice has *Pnma* symmetry arising from orthorhombic distortion.(*23, 24*) The symmetry-breaking introduced by this distortion has important consequences for the optical properties of the emitters.

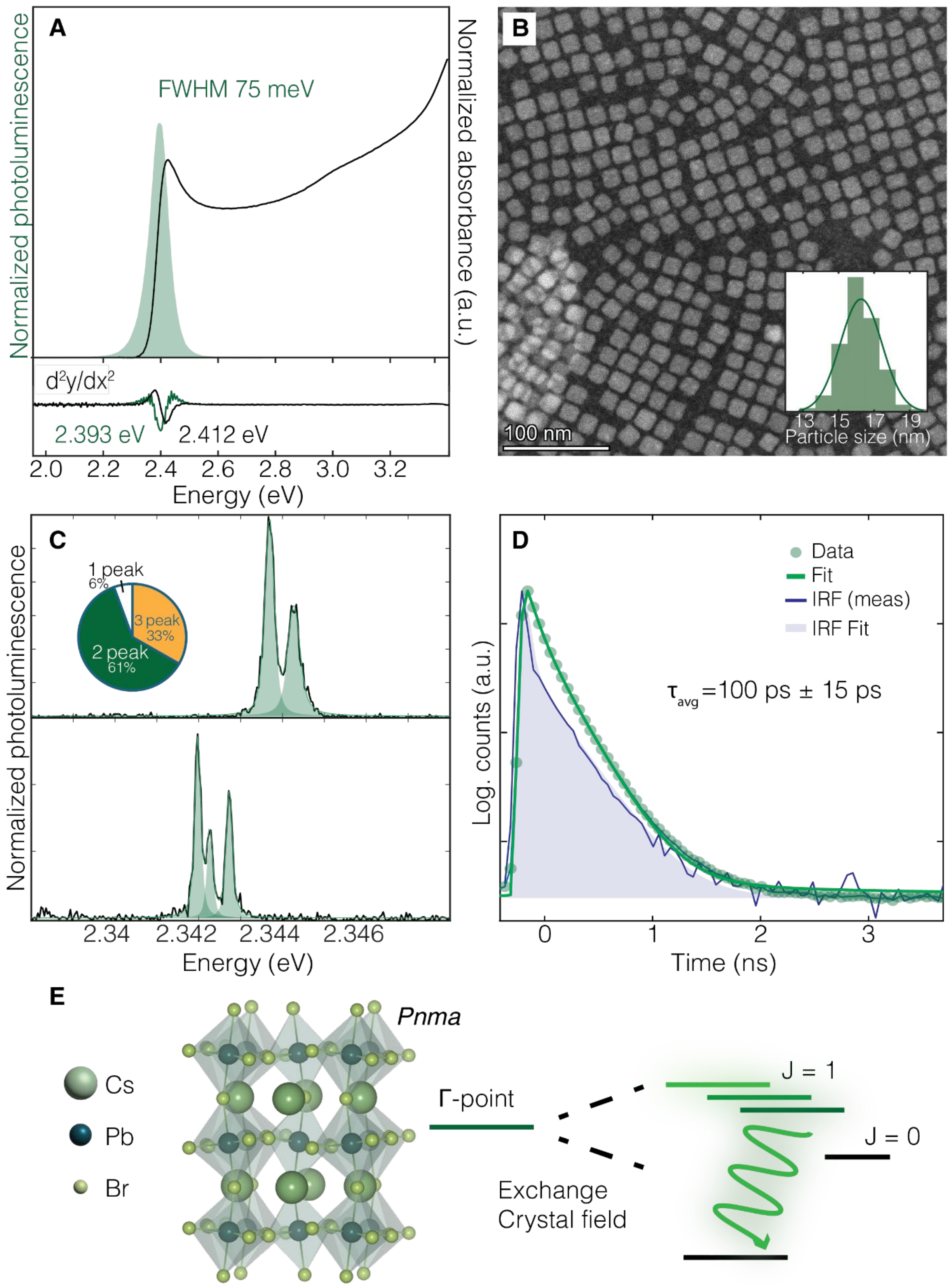

**Figure 1. Optical and structural characterization of as-synthesized $CsPbBr_3$ nanocrystals (NCs).** (**A**) Room temperature ensemble absorption and photoluminescence (PL) spectra with second derivative analysis showing an absorption peak at 2.412 eV and PL emission at 2.393 eV, with a full width at half maximum of 75 meV. (B) Dark field transmission electron micrograph of the NCs. From electron microscopy image analysis, the size varies by imaged area: the subsection shown here has an average size of 16.8 nm ± 0.1 nm. (C) 4 K PL spectra of single nanocrystals show multiple peaks. Two thirds of the studied NCs show 2 peaks while one third have three peaks. (D) A representative time resolved single photon counting trace. The average PL lifetime across the studied NCs is 100 ps ± 15 ps. (E) The NC lattice consists of corner-sharing Pb-Br octahedra bonded by Cs+ cations. The symmetry group, Pnma, supports a bandgap at the Γ-point of the Brillouin zone. Intrinsic asymmetries cause splitting of the lowest exciton state into a manifold: an optically bright triplet and dark singlet.

### Single nanocrystal spectroscopy

We turn to low-temperature optical spectroscopy of single particles to uncover the intrinsic physics obscured by the ensemble. For a set of 20 NCs, we measured single-particle photoluminescence (PL) spectra and PL lifetimes (Fig. 1C,D). Individual NC spectra show between one and three peaks. Each spectral peak is fit by a Voigt profile representing the natural (Lorentzian) linewidth broadened by a Gaussian, yielding an average full width at half maximum of 280 ± 100 μeV across the sample. The Gaussian contribution could arise from spectral diffusion, elastic phonon scattering, or limitations of the spectrometer resolution (150 μeV). In addition, the NCs exhibit fast radiative lifetimes of 100 ps ± 15 ps, approaching our multiexponential instrument response function with short (long) component of 15 ps (230 ps) (Fig. 1E, fig. S4).

In orthorhombic phase $CsPbBr_3$, the conduction and valence bands are derived from the Pb-Br octahedral orbitals at the Γ-point of the Brillouin zone.(25) In the transition from the cubic to the orthorhombic crystal structure, the band edges change from the R to the Γ point due to zone folding resulting from the unit cell expansion of the orthorhombic crystal structure.(*26, 27*) A Bloch exciton formed from an electron and hole at the band edge has fine structure resulting from electron-hole exchange and crystal-field interactions (Fig. 1E).(*28*) This yields a set of closely spaced exciton fine structure states: a lower energy dark singlet, J = 0, and an upper energy bright triplet state, $J_z = (+1, 0, -1)$. The fine structure governs the selection rules for photon emission: the singlet is optically dark whereas the spin-allowed triplet is bright, explaining the triplet peaks observed in the single NC spectra.(*29*) In addition to selection rules, the rapid emission of 100 ± 15 ps is explained by collective oscillator superradiance.(*30, 31*)

Symmetry reduction has direct consequences for the polarization of the emitted photons. In a perfectly symmetric cubic lattice, a degenerate exciton manifold emits right- and left-circularly polarized photons. From the lower symmetry orthorhombic lattice, the emitted light is linearly polarized (Fig. 2A). Therefore, photons emitted from the lowest-lying exciton triplet state serve as reporters of the underlying Bloch wavefunctions.

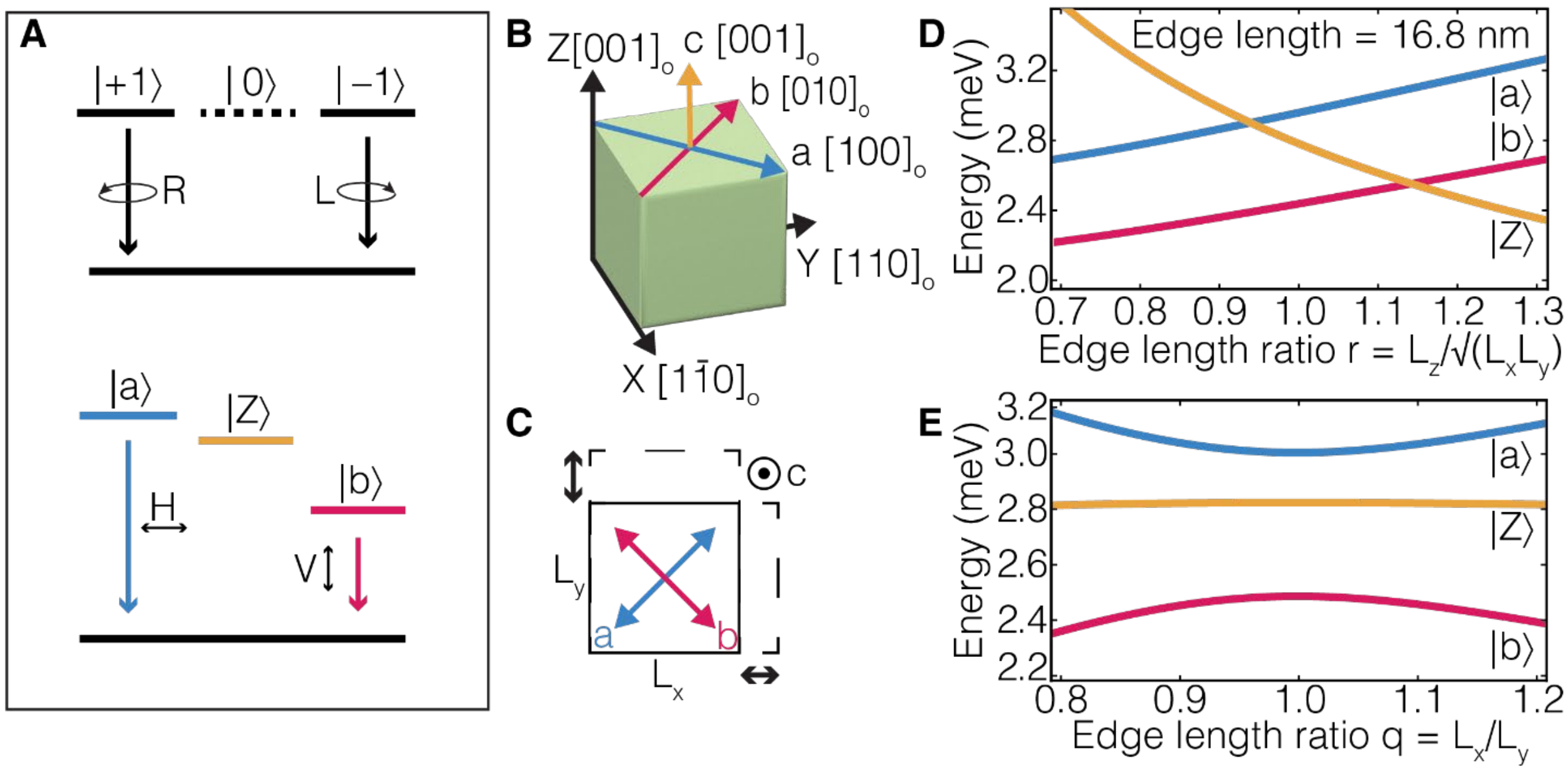

**Figure 2. The triplet manifold supports polarized photon emission.** (A) A cubic lattice with a degenerate triplet would have two optically active states coupling to oppositely-handed circularly polarized light. With the asymmetry of the underlying *Pnma* Bloch functions and electron-hole exchange interaction, the triplet state splits into three distinct, linearly-polarized fine structure levels. The sketch shows the case where the $|Z\rangle$ state transition dipole is aligned parallel to the optical axis. (B) The orientation of the transition dipoles of the triplet levels overlaid on a cartoon cube-shaped NC. Whereas the underlying lattice is orthorhombic, with lattice vectors denoted by **a**, **b**, and **c**, the nanocrystal faces correspond to the pseudocubic planes and are normal to the directions X, Y, and Z. (C) View down the Z-axis of the nanocube. In a perfect cube-shaped nanocrystal, the three triplet states would have transition dipoles aligned to the orthorhombic symmetry axes, **a**, **b**, and **c**. Deviations in the ratio of the lengths of the axes, denoted by $L_x$ and $L_y$, affect the relative energies of the triplet states. If $L_x \neq L_y$, the excitons with transition dipoles aligned to **a** and **b** in the absence of long-range exchange couple and mix, causing an anticrossing as the ratio $L_y/L_x$ deviates from 1. Panels (D) and (E) show the consequence of volume-preserving variations of the Z dimension and the ratio of the X/Y dimensions on the energies of the $|a\rangle$, $|b\rangle$, and $|Z\rangle$ levels, respectively.

Given the *Pnma* crystal structure and the presence of up to three spectrally distinct peaks, we expect linearly polarized emission from the LHP NCs. The NC are near-cube-shaped with facets normal to the pseudocubic lattice vectors, while the underlying crystal lattice is orthorhombic. Consequently, for near cube-shaped NCs, two of the exciton fine structure levels, $|a\rangle$ and $|b\rangle$, have transition dipoles aligned to orthorhombic lattice vectors ***a*** and ***b*** which point to the NC corners (Fig. 2B).The third level, $|Z\rangle$, has its transition dipole aligned to orthorhombic lattice vector ***c***.(*32*) The particle shape is critical as long-range exchange interactions depend sensitively on edge symmetry and roundedness (fig. S5).(*32*) Calculations for our particle size using the model of Ref. 32 indicate that the ordering of the triplet state energies depends on whether the nanocrystal is slightly oblate or prolate relative to a perfect cube (Fig. 2C, see SM). We expect the $|Z\rangle$ exciton state to lie intermediate between the $|a\rangle$ and $|b\rangle$ exciton states for the average nanocrystal size and overall cubic shape studied here (Fig. 2D, E).

### Polarization spectroscopy

To obtain the polarization state of each fine structure peak, we record the emission spectrum as a function of quarter-wave plate angle relative to a linear polarizer and fit the individual spectral peaks with Voigt profiles (fig. S6-7). Representative fits for NCs with two and three peaks are shown in Fig. 3A, and more in fig. S8. The integrated area of each peak modulates with waveplate angle with a period of $\frac{\pi}{2}$ (Fig. 3B), with different peaks oscillating $\frac{\pi}{4}$ out of phase. Fitting to the functional form of Equation S1.1 yields the full Stokes vector ($S_0$, $S_1$, $S_2$, $S_3$) of each spectrally-resolved peak.

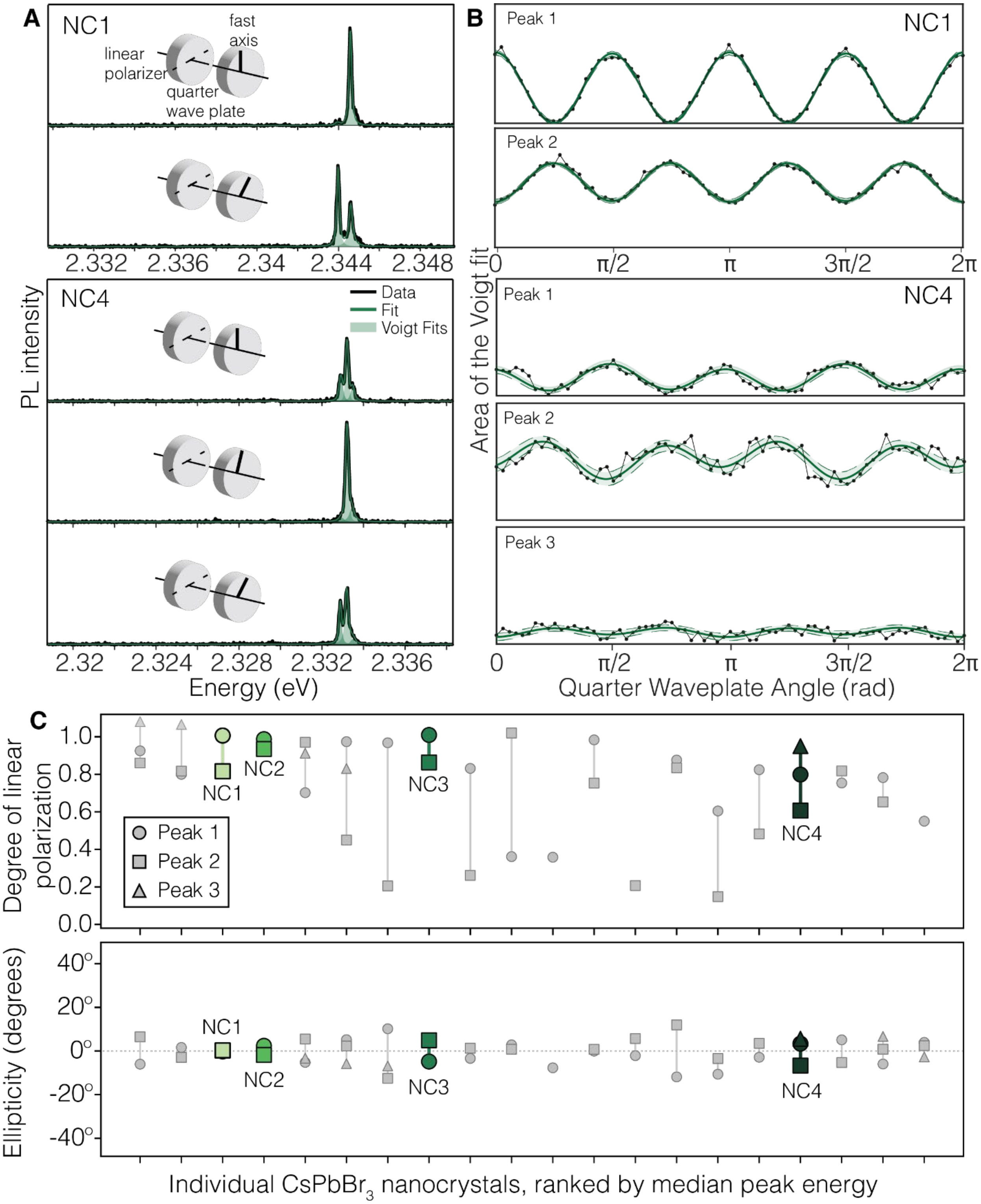

**Figure 3. Polarization analysis of single NC PL spectra.** (**A**) shows the as-measured PL spectra for two representative NCs with varying angles of the rotating quarter waveplate analyzer (inset). NC1 has two visible peaks while NC4 has three visible peaks. Each underlying peak is fit to a Voigt profile and the sum fit is overlaid on the data. (B) Integrated areas of the Voigt fits for each resolved peak in Panel (A) as a function of quarter waveplate angle. (C) Fitting the traces in Panel (b) yields the Stokes parameters, which are used to determine the total degree of linear polarization (upper panel) and ellipticity (lower panel) across 20 measured NCs.

Figure 3C summarizes the degree of linear polarization (DOLP $= \frac{\sqrt{S_1^2+S_2^2}}{S_0}$) for every resolved peak across the 20 NCs studied. Fig. 3D shows the corresponding ellipticity angles; the ellipticity angle, $\chi = \frac{1}{2}\arctan\left(\frac{S_3}{\sqrt{S_1^2+S_2^2}}\right)$, illustrates the circular character of the polarized component of the emission: $\chi = 0°$ corresponds to purely linear and $\chi = \pm 45°$ to purely circular polarization. The highlighted NCs, labeled NC1-4, are those that are analyzed further in the text.

The apparent DOLP varies substantially between NCs, indicating their varying orientation relative to the optical axis. For a cuboidal NC resting flat on the substrate, if $[001]_o$ coincides with the optical axis, the $|a\rangle$ and $|b\rangle$ exciton transition dipoles lie in the sample plane and are detected as two orthogonal, linearly polarized peaks. The $|Z\rangle$ exciton transition dipole is oriented along the collection axis and radiates predominantly outside the collection cone. In this upright geometry, the $|Z\rangle$ state contributes a weak nonpolarized intensity across all polarizer angles due to our high numerical aperture objective (NA = 0.9). Spectrally, it is too weak to be resolvable. Example NC1, with only two visible spectral peaks, corresponds to this orientation (Fig. 3A, upper panel).

As the NC tilts in the collection plane, the $|Z\rangle$ state transition dipole projects into the plane and becomes visible as a third spectral peak (Fig. 3A, lower panel), while the $|a\rangle$ and $|b\rangle$ exciton transition dipoles lose in-plane character and appear partially depolarized due to a larger component in the high-NA regime; their apparent DOLP therefore decreases with increasing tilt. In contrast, the ellipticity angles are insensitive to tilt (Fig. 3D) and cluster near zero (median $|\chi|$ = $4° \pm 6°$). Each pair is equal and opposite in ellipticity and orthogonal at 90° (fig. S8), except in the three-peak cases, where projection into the plane destroys the orthogonality.

**Exciton fine structure relaxation**

Having established the identity and orientation of the emitting states, we now turn to relaxation between them. We measured NC1-4 as highlighted in the previous section: three exhibiting two spectral peaks and one exhibiting three peaks. EFSR between linearly polarized fine structure states can be quantified using polarization-resolved second-order intensity correlations,

$$g^{(2)}_{\theta\theta'}(\tau) = \frac{\langle I_\theta(0) I_{\theta'}(\tau)\rangle}{\langle I_\theta(0)\rangle\langle I_{\theta'}(\tau)\rangle}$$

which track how the polarization relationship between two photons persists as a function of their emission time delay, $\tau$, in polarization channels, $\theta(\theta')$. We measure $g^{(2)}_{\theta\theta'}(\tau)$ using a Hanbury Brown-Twiss interferometer equipped with an independent polarization analyzer in each arm (Fig. 4A, fig. S9). From it, we define the degree of correlation (DOC),

$$DOC = \frac{g^{(2)}_{\mu\mu}(0) - g^{(2)}_{\mu\mu'}(0)}{g^{(2)}_{\mu\mu}(0) + g^{(2)}_{\mu\mu'}(0)}$$

where $\mu$ and $\mu'$ denote two orthogonal polarization bases (e.g., H and V). The degree of correlation ranges from +1 to −1: +1 is perfect correlation, −1 perfect anticorrelation, and 0 no polarization correlation.(*16*)

The system is initialized in a biexciton (BX) state, a bound state of two excitons, by nonresonant excitation, and it relaxes sequentially through the biexciton-exciton (BX-X) cascade, conserving total angular momentum (Fig. 4B). Absent EFSR, decay through exciton state $|a\rangle$ emits an H-polarized BX photon followed by an H-polarized X photon, while decay through exciton state $|b\rangle$ emits a V-polarized BX photon followed by a V-polarized X photon. When it proceeds through $|a\rangle$, the co-polarized correlation $g_{HH}^{(2)}(0)$ monitors the conditional probability that the exciton decays through its initial path, whereas the cross-polarized correlation $g_{HV}^{(2)}(0)$ measures inter-state population transfer, or EFSR, denoted by $\gamma_{ab}$ and $\gamma_{az}$ in Fig. 4B.

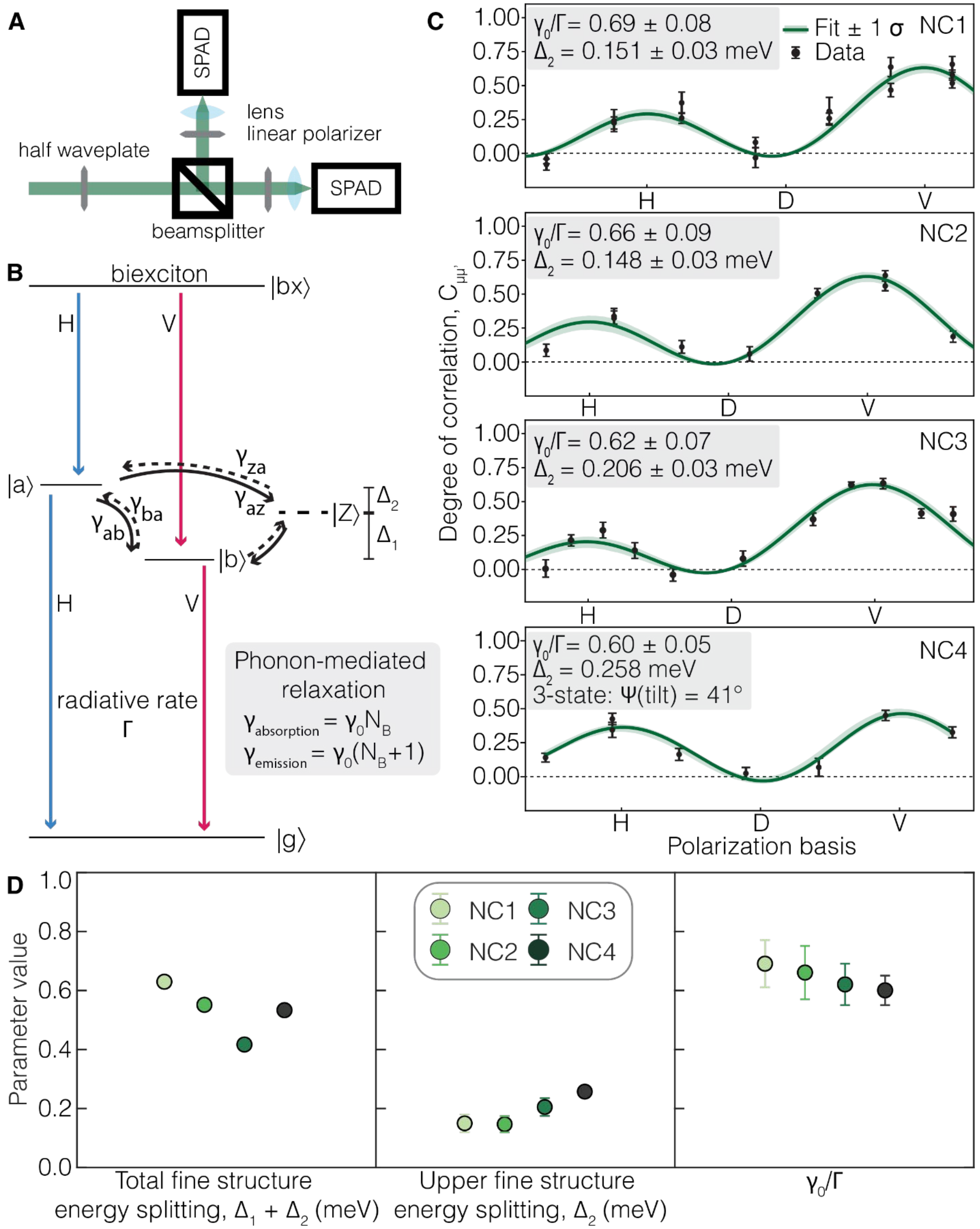

**Figure 4. Dynamic interactions between fine structure states are uncovered through second-order polarization correlations.** (**A**) The setup. A half waveplate before a Hanbury-Brown-Twiss interferometer is used to rotate the measurement basis. Linear polarizers before single photon avalanche diodes select co- and cross-polarized orientations. (B) The underlying model of population transfer in $CsPbBr_3$ NCs. Radiative rates are shown in colour, while nonradiative rates are shown in black and white; the exciton radiative rate constant is denoted by Γ. A biexciton

state conserves angular momentum upon decay through the exciton state and both photons out of the biexciton cascade share the same linear polarization (H or V); however, momentum can be nonradiatively transferred to an energy-matched acoustic phonon, labeled $\Delta_1$ and $\Delta_2$. EFSR pathways out of state $|a\rangle$ are labeled $\gamma_{ab}$ and $\gamma_{az}$, while $\gamma_{ba}$ and $\gamma_{za}$ indicate leakage into state $|a\rangle$. Dashed lines denote the absorption of a phonon; full lines, emission. Assuming microscopic reversibility, phonon absorption (emission) are temperature-dependent via the Bose-Einstein distribution, $N_B(\Delta, T)$, and zero-temperature phonon coupling rate constant, $\gamma_0$. (C) Polarized second order correlations track relaxation through the cascade, with cross-polarized correlation indicating EFSR. Extracted fit parameters are shown in the insets. (D) A common zero-temperature coupling strength $\gamma_0/\Gamma$ of 0.60 $\pm$ 0.05 indicates a universal phonon-mediated EFSR mechanism across the studied NCs.

Measuring the DOC as a function of polarization basis angle tracks EFSR directly and separates it from pure dephasing. Pure dephasing leaves populations unaffected and therefore does not affect the DOC, while EFSR redistributes population between fine-structure states and therefore does. In the absence of EFSR, an exciton decaying through one bright state remains there for the full exciton lifetime: the co-polarized correlation would show strong bunching at zero time, the cross-polarized correlation would show antibunching, and the DOC would approach +1. EFSR introduces transitions between the fine-structure states on the timescale of the exciton lifetime, so population leaks into the orthogonal state with some probability. This reduces the co-polarized correlation, raises the cross-polarized correlation at zero time, and lowers the DOC.

Figure 4C shows the DOC versus polarization basis for four LHP NCs. The DOC oscillates with analyzer angle, peaking when the basis aligns with intrinsic polarization axes (H/V) and dropping to a minimum at the diagonal. The first three NCs behave as on-axis emitters: their H and V channels show different correlation amplitudes and the V path reaches a DOC above 0.6, higher than reported for some epitaxial QDs.(*10*) The fourth NC is tilted off-axis. The tilt rotates the |Z⟩ state transition dipole into the detection plane and mixes all three orthogonal states in the linear-polarized basis, so the H and V correlation amplitudes become nearly equal.

The observed behavior is described by a model originally developed by Das and Agarwal for a two-state system.(*16*) We extend the model to the full three-state triplet manifold to extract the EFSR rate constant (see SM). For the three on-axis NCs, the $|Z\rangle$ state transition dipole lies along the crystal axis that is parallel to the collection axis, and we fit the energy gap $\Delta_2$ (Fig. 4C). For the tilted NC, we directly measure $\Delta_1$ and $\Delta_2$, and we fit the tilt angle $\Psi_{tilt}$ that projects the $|Z\rangle$ state transition dipole into the $|a\rangle$-$|b\rangle$ plane. Kinetically, $|Z\rangle$ remains coupled to $|a\rangle$ and $|b\rangle$ through phonon-mediated relaxation. The phonon occupation follows the Bose–Einstein distribution, $N_B = \left(e^{\frac{\Delta}{k_B T}} - 1\right)^{-1}$, where $\Delta$ is the energy splitting between each of the fine structure levels. Further, the higher-lying exciton state can relax through spontaneous emission of a phonon, even at zero temperature.(*33*) The EFSR rates between fine structure levels take the form,

$$\gamma_{absorption} = \gamma_0 N_B$$

$$\gamma_{emission} = \gamma_0 (N_B + 1)$$

where $\gamma_0$ is a rate constant setting the zero-temperature phonon coupling strength. The extra term in $\gamma_{emission}$ reflects the contribution from spontaneous phonon emission.

The extracted parameters fall in a narrow range across the four independent NCs (Fig. 4D). For NC1-3, the fit outputs the $|Z\rangle$-state position and the normalized EFSR rate. The $|Z\rangle$ state lies 150-250 μeV below the upper fine structure state, with EFSR between 0.6 and 0.7 of the radiative rate. For NC4, which has three spectral peaks, the directly measured $\Delta_2$ splitting is 260 $\mu eV$, consistent with the fit values of $\Delta_2$ for NC1-3. The normalized EFSR coupling constant sits at the lower end of the NC1-3 values at $\gamma_0/\Gamma = 0.60 \pm 0.05$. Across all NCs, the average normalized EFSR coupling constant is $0.64 \pm 0.04$, about two-thirds of the radiative rate constant of the emitters. The asymmetry seen in the DOC of NC1-3 follows (Fig. 4C). Downhill relaxation includes spontaneous phonon emission, so it survives as $T \rightarrow 0$, whereas the uphill rate scales with the phonon occupation and vanishes. The lower-energy bright state is therefore protected at zero temperature, where its degree of correlation approaches unity, while the higher-energy state relaxes downward regardless of temperature. The presence of the intermediate $|Z\rangle$ state lowers the barrier of phonon coupling and facilitates effective population transfer at 4 K.

Efficient phonon coupling drives EFSR in LHP NCs. The measured phonon energies sit an order of magnitude below the lowest reported optical phonons in $CsPbBr_3$.(*34, 35*) Acoustic modes in this range (150-200 μeV) have been resolved in related LHP nanomaterials,(*36, 37*) but their importance in $CsPbBr_3$ has been de-emphasized. To further investigate, we model $CsPbBr_3$ NCs of average edge length 16-20 nm as a free cube using the elastic tensor for orthorhombic $CsPbBr_3$ and estimate the available acoustic mode energies (fig. S10).(*38-40*) A set of acoustic modes spans the measured triplet fine structure splitting, with the lowest-energy mode set by particle size. For a perfect cube, this mode sits at 100-120 μeV across the size range shown, matching the measured fine structure splitting. The lowest-lying modes are volume-conserving shear modes, exactly those that would couple orthogonal states.

An EFSR coupling strength on the order of the radiative rate requires strong coupling to phonon modes, most likely to the shear acoustic modes identified above. Deformation potential coupling is intrinsically weak for shear strain, and its coupling strength scales with wavevector as $q^{1/2}$ *versus* $q^{-1/2}$ for piezoelectric potential; piezoelectric coupling therefore dominates at the small wavevectors of lowest-energy confined modes.(*41, 42*) Piezoelectric coupling is formally forbidden in the centrosymmetric *Pnma* structure assigned to the bulk crystal, but local fluctuations could offset the octahedra from the cations to create transient polarity within the NC.(*43*) A strain-stabilized monoclinic polar phase for $CsPbBr_3$ is predicted and would make piezoelectric coupling directly allowed.(*44*)

Phonon-mediated EFSR is detrimental to the coherence times of LHP NCs. The coherence time of the upper state has a ceiling set by spontaneous phonon emission even at zero temperature: a zero-temperature EFSR coupling strength $\gamma_0 = 0.64\ \Gamma$ shortens its population lifetime by a factor of 1.6 and limits its coherence to roughly 0.6 of the radiative transform limit. Equivalently, this transfer removes about 40% of excitons prepared in the upper state before they radiate. The lower state is protected at zero temperature, but rapidly loses coherence once phonons are thermally populated and can be absorbed. The amount of coherence loss depends on the thermal availability of phonon modes matching the fine structure splitting. In the best measured nanoparticle at 4 K (NC1) population transfer from EFSR alone limits coherence to $0.81 \pm 0.03$ of the transform limit; in NC3, the smaller splitting drastically enhances the transfer out of the state, leading to a maximum coherence of $0.58 \pm 0.07$ of the transform limit (table S1).

EFSR can be mitigated extrinsically, through device and field integration, or intrinsically, through chemical modification of the nanocrystal. The radiative rate can be Purcell-enhanced to outcompete EFSR,(*3, 45, 46*) and fine structure splitting can be tuned in a fixed NC via electromagnetic fields.(*47-49*) Resonant, polarization-selective excitation into the lowest fine structure state can prepare the exciton in its most coherent state, though this gain appears only at ultra-low temperatures, where phonon absorption freezes out. Intrinsically, minimizing the fine structure splitting is most effective: lowering it below the acoustic phonon modes eliminates phonon coupling, and lowering it further, below the emitters' natural linewidth, erases which-path information for polarization-entangled photon pairs.(*10, 15, 50*) Morphology tunes both the acoustic modes and the splitting: rounder NCs have stiffer acoustic modes (fig. S10).(*40*) Because the splitting is governed by electron-hole exchange, it also depends on NC size and shape.(*32, 51*) Thus, NC size and morphology influence both the radiative rate and the energy splitting. Anisotropic morphologies, such as nanoplatelets and nanowires, may provide a practical route to fast emission while suppressing fine-structure splitting.(*52, 53*)

We follow relaxation within the bright triplet manifold directly in single $CsPbBr_3$ NCs at 4 K, and identify EFSR as a major, previously overlooked channel of decoherence, intrinsic to the nanocrystal rather than its environment. This mechanism can be reduced through established synthetic control over nanocrystal size, shape, and fine structure splitting, giving LHPs a defined route to higher-coherence photon sources without sacrificing the scalable, solution-based production that distinguishes them from epitaxial and atomic emitters.

**Acknowledgements**

TŠ and NLB acknowledge support from the Department of Energy, Office of Basic Sciences grant DE-SC0021650. CJK acknowledges support from the U.S. Army Research Office through the Institute for Soldier Nanotechnologies (Award No. W911NF2320121). ALE acknowledges the support of the Office of Naval Research. Part of this work was authored by the Alliance for Sustainable Energy, LLC, the manager and operator of the National Laboratory of the Rockies for DOE under contract no. DE-AC36-08GO28308 (PS). The views expressed in the article do not necessarily represent the views of the DOE or the US Government. Calculations of the exciton fine structure in $CsPbBr_3$ nanocrystals were supported as part of the Center for Hybrid Organic–Inorganic Semiconductors for Energy (CHOISE), an Energy Frontier Research Center funded by the Office of Basic Energy Sciences, Office of Science in the US Department of Energy. TŠ would additionally like to acknowledge a Natural Sciences and Research Council of Canada Graduate Scholarship. AJJ was supported by a National Science Foundation Graduate Research Fellowship. ST acknowledges the European Union's Horizon Europe research and innovation program under the Marie Skłodowska-Curie Funding Program (Project SUPER-QD, Grant Agreement No.101148934). This work was carried out in part through the use of MIT.nano and MIT Materials Research Lab facilities. TŠ would like to thank Oliver Nix and Kaelyn McFarlane-Connelly for experimental assistance and advice, and Felix Knollman for productive conversations.

**Author Contributions:**

Conceptualization: TŠ, AEK, PCS, ALE, MGB
Methodology: TŠ, AJJ, NB, CJK, AEK, ST, PCS
Software: TŠ, NB
Formal Analysis: TŠ, AJJ
Investigation: TŠ, AJJ, CJK
Visualization: TŠ, AJJ, PCS, CJK, ST
Funding acquisition: TŠ, AEK, MGB
Supervision: MGB
Writing – original draft: TŠ
Writing – review & editing: TŠ, AJJ, PCS, ALE, MGB

**Competing Interests:** Authors declare that there are no competing interests.

**Data, Code, and Materials Availability:** Data, code, and materials are available upon request.

**List of Supplementary Materials:**

Materials and Methods
Supplementary Text 1: Experimental and fitting
Supplementary Text 2: Exciton fine structure model
Fig S1-10
Table S1 and S2

# Supplementary Materials for

## Exciton Coherence in $CsPbBr_3$ Nanocrystals is Bounded by Phonon-Mediated Bright-Triplet Relaxation

Tara Šverko, Annette J. Jones, Chantalle J. Krajewska, Peter C. Sercel, Alexander L. Efros, Alexander E. Kaplan, Niamh L. Brown, Stefano Toso, Moungi G. Bawendi*

Corresponding author: mgb@mit.edu

**The PDF file includes:**

## Materials and Methods

Materials

1-octadecene (90%, Sigma-Aldrich), lead acetate trihydrate ($Pb(OAc)_2 \cdot 3H_2O$) (99.999%, Sigma-Aldrich), cesium acetate (99.9%, Sigma-Aldrich), oleic acid (90%, Alfa Aesar), bromotrimethylsilane (97%, Sigma-Aldrich), 3-(*N*,*N*-dimethyloctadecylammonio)propanesulfonate (ZwL) (>99%, Sigma-Aldrich), lead dibromide (anhydrobeads, 99.999% trace metals basis (perovskite grade), Sigma-Aldrich), toluene (anhydrous, 99.8%, Sigma-Aldrich), ethyl acetate (EtOAc) (anhydrous, 99.8%, Sigma-Aldrich), poly(methyl methacrylate) (PMMA) (average molecular weight 120kD, Sigma-Aldrich) were used without further purification.

Methods

*Hot injection synthesis of ~18 nm edge length $CsPbBr_3$ nanocrystals (NCs)*

10 mL-octadecene, 1 mL oleic acid, 0.2 mmol $Pb(OAc)_2 \cdot 3H_2O$, 0.15 mmol Cs(OAc) and 0.1 mmol ZwL were loaded into a 50 mL round bottom flask and degassed at 120°C for 1 hour while stirring. The hot injection was done by rapidly heating the solution under nitrogen gas to 240°C, whereupon 0.6 mmol of the bromotrimethylsilane precursor was rapidly injected with a syringe. The reaction flask was immediately placed in an ice bath and cooled to room temperature and the entire contents were centrifuged at 14310g for 10 minutes. The pellet was redispersed in 6 mL toluene, followed by three washing steps with EtOAc (the ratio of toluene: EtOAc was 1:2). The final product was then obtained and kept dispersed in toluene; any larger or unstable particles were removed via final centrifugation with no antisolvent and the pellet was discarded.

*Ensemble room temperature absorbance and emission*

The sample for room temperature absorbance and emission spectra was prepared by diluting the stock 100-fold in toluene. Absorption measurements were taken on an Agilent Technologies Cary 5000 UV-Vis-NIR spectrometer. Photoluminescence spectra were taken on a HORIBA Jobin Yvon Fluoromax 3 spectrofluorometer.

*Dilute single perovskite NC sample preparation*

Dilute films of $CsPbBr_3$ nanocubes were prepared for single particle optical analysis at low temperature. A 5% solution of 120 kD poly(methyl methacrylate) in toluene was prepared under inert atmosphere and stored over molecular sieves (4A) for over 24 hours. The solution was saturated with the native zwitterionic ligands and an added amount of lead dibromide, after which it was filtered through a 0.2 um PTFE filter. All chemicals, vials, and pipette tips were kept under inert atmosphere or otherwise dried prior to use. The nanocubes were diluted in this solution sequentially, up to a 50000-fold dilution.

Additionally, quartz substrates (z-cut, optically polished on one side) were cleaned using Hellmanex, after which the surface was prepared via silanization with octadecyltrimethoxysilane (OTMS). A 1% v/v solution of OTMS in toluene was prepared, and the substrates were submerged within. 0.5% v/v butylamine catalyst was added, and the substrates were left to react for 4 hours under nitrogen flow. The substrates were then removed and left on a hot plate under vacuum overnight to remove trace solvent. The substrates were sonicated in clean, dry toluene to remove excess reagent. The protocol was repeated to create a more perfect hydrophobic monolayer.

The nanocubes were deposited in an even, thin layer on the clean, treated quartz substrates by spin coating. 20 uL of the dilute solution were statically deposited, after which the spin rate was ramped up to 500 rpm for 10 seconds, then 1500 rpm for 1 minute, and 3000 rpm for another minute.

*X-ray Diffraction*

X-ray diffractograms were acquired on a multipurpose diffractometer (PANalytical X'Pert Pro) using a 1.8 kW Cu-K$\alpha$ X-ray source in a Bragg-Brentano geometry. The sample was prepared by dropcasting a concentrated solution of perovskite NCs onto a zero-diffraction plate (MTI Corp, silicon crystal, P-type B-doped).

*Transmission Electron Microscopy*

Dark field transmission electron micrographs (TEMs) and high angle annular dark field scanning TEMs were acquired on a probe-corrected Thermo Fisher Themis Z G3 operated at 200 kV and collected on Super-X detectors. Films of perovskite NCs were dropcast onto a copper grid (400 mesh) with an amorphous carbon grid (Ted Pella).

*Low temperature confocal spectroscopy*

Single nanocrystal experiments at 4 K were done on a homebuilt setup. The sample was loaded into a Montana Instruments Cryostation s50 closed-cycle liquid helium cryostat with a cryo-optic objective (Zeiss EpiPlan Fluor 100× 0.9 NA). The sample was excited at 490 nm excitation wavelength (Toptica Photonics FemtoFiber Pro) with a pulse width of 220 fs and a repetition rate of 80 MHz. Using polarization-maintaining mirrors (Edmund Optics TECHSPEC Polarization Maintaining Mirrors, 532 nm), the sample light was guided through a 10:90 R:T nonpolarizing beamsplitter (Thorlabs) and filtered by a 500 nm 0-degree longpass filter (Thorlabs). After a 1:1 telescoping pinhole (achromatic lenses, 80 mm focal length, 50 μm pinhole) and a zero-order broadband half-wave plate, the sample light was guided either to a spectrometer or to a Hanbury Brown-Twiss interferometer. The spectrometer, a Princeton Instruments Acton SP500i with a holographic 2400 g/mm grating and Princeton Instruments ProEM CCD cooled array, was used with a broadband zero-order rotating quarter-wave plate and linear polarizer to measure energy-resolved Stokes parameters. On the other arm, a Hanbury Brown-Twiss setup was constructed using a 50:50 nonpolarizing beamsplitter (Thorlabs) and equipped with two linear polarizers before reaching single photon avalanche diodes, each with an individual resolution below 30 ps (MPD). Data was collected on a Swabian Instruments Timetagger Ultra with 8 ps rms jitter.

**Supplementary Text 1: Experiment and Fitting**

<u>Size and shape analysis of as-synthesized $CsPbBr_3$ NCs</u>

We analyzed the ensemble-averaged size and distribution through XRD and STEM analysis. Figure S1 shows extended STEM micrographs. The XRD analysis shows that the sample is composed of CsPbBr3 nanoparticles with no noticeable impurities (fig. S2). The pattern can be indexed according to the (pseudo)cubic phase, as reflections that would identify it as orthorhombic are too weak to be detected (this does not mean that the structure is cubic). The pseudocubic lattice constant is approx 5.85 Å, slightly smaller than expected for nano-$CsPbBr_3$, potentially a sample alignment issue. The dropcast sample produced an alteration of the peak intensities as the (faceted) particles oriented preferentially with their (100) cubic facets parallel to the substrate. However, the orientation is not perfect and there is no sign of superlattice interference, which points to rounded cubes. The Williamson-Hall plot analysis of XRD peak linewidth (fig. S3) indicates an average NC size of 17.8 nm, checked by simulating the expected profile for the (100) peak (fig. S3 inset).
Further analysis of size and shape was done using the darkfield TEM and HAADF STEM images. A mask was applied to the images and the particles were isolated from the dark background, shown for the main text HAADF STEM image (fig. S5A,B). Figure S5(C-E) show the extracted average edge length, squareness, and roundedness along with 1-sigma bounds. Across the STEM and TEM images, the average particle edge length is around 17 nm, with a distribution of $\pm 1$ nm at 1-sigma, agreeing well with the XRD analysis. The particles additionally support the fact that the edges are slightly rounded. The particles are largely square (projection of cubic), with a truncated Gaussian showing a mean around 1, meaning no significant edge elongation away from a square profile.

Time-resolved photoluminescence (PL) fitting
PL traces were histogrammed to determine on and off states (photoluminescence intermittency and charging). The on-state PL lifetimes were fit via numerical reconvolution. The instrument IRF was recorded at 520 nm and used to fit each PL trace to a monoexponential. The polymer background scatter was also recorded and added as a constant parameter to the fit. The data was normalized before fitting.

$$N\left[IRF(t)\otimes e^{-\left(\frac{t}{\tau_{QD}}\right)}\right]+BG \qquad (S1.1)$$

A sample fit for NC 1 in the main text is shown in fig. S4(A). Panel (B) shows the average extracted PL lifetimes for each measured NC, with the dots highlighted in the main text also highlighted for comparison. Panel (C) finds the ensemble averaged single NC lifetime of this sample: 100 ps $\pm$ 15 ps.

Depolarization of the confocal microscope
The confocal microscope was carefully calibrated to remove polarization artifacts. The confocal microscope is shown in fig. S6. Path A leads to the Hanbury Brown Twiss interferometer, and both the transmission and reflection beams were calibrated. Path B leads to the spectrometer. Linear and circularly polarized light at 532 nm (2.331 eV) was passed through the setup and characterized by Stokes analysis (fig. S7). The analysis shows that circularly polarized light retained 91%, 87%, and 99% of its character for Path A reflection, Path A transmission, and Path B, respectively. Linearly polarized light retained 81%, 79%, and 88% of its character, respectively. The remaining loss was largely to general depolarization and not retardance or birefringence, as shown by the overall degree of polarization.

Stokes analysis of polarized light: The rotating quarter waveplate
The polarization state of a stream of light is well described by its Stokes parameters, $S0$, $S1$, $S2$, and $S3$. $S0$ encompasses the total intensity of the light, $S1$ the difference between horizontal and vertical components, $S2$ the difference between diagonal components, and $S3$ the difference between right-hand and left-hand circular components. One method to measure the Stokes parameters of light is by recording its intensity as it passes through a rotating quarter waveplate (QWP) followed by a linear polarizer (LP). As the QWP rotates the intensity varies as

$$I(\theta) = 12(S0 + S1cos2(2\theta) + S2\sin(2\theta)\cos(2\theta) + S3\sin(2\theta)) \qquad (S1.2)$$

Where $\theta$ is the angle between the QWP fast axis and the LP transmission axis. In this experiment, to obtain the Stokes parameters of each FSS, after passing through the QWP and LP the emission is spectrally separated by a 2400 g/mm grating and collected on a CCD camera with one second integration time. Between each measurement the QWP is rotated by 5 degrees on a digital rotation mount. The intensity of each resolved fine structure state is then fit to a Voigt profile. For each peak the Stokes parameters are extracted by rewriting the expression for the intensity in Fourier terms. This yields

$$I(\theta) = 12\big(A + Bsin(2\theta) + Ccos(4\theta) + Dsin(4\theta)\big) \qquad (S1.3)$$

where $A$=$S0$+$S12$, $B$=$S3$, $C$=$S12$, $D$=$S22$. For a measurement consisting of $N$ equally spaced QWP positions the Fourier terms are

$$A = 2N\sum InNn = 1$$
$$B = 4N\sum Insin(2\theta n)Nn = 1$$
$$C = 4N\sum Incos(4\theta n)Nn = 1$$
$$D = 4N\sum Insin(4\theta n)Nn = 1 \qquad (S1.4)$$

Once obtained, the Fourier terms are then converted back to the Stokes parameters via the relations $S0$=$A$−$C$, $S1$=2$C$, $S2$=2$D$, $S3$=$B$.(53)

Additional Stokes analysis of single $CsPbBr_3$ nanocrystals

Additional representative spectra, Voigt fits, and area under the Voigt fits as a function of quarter waveplate are shown in fig. S8 A and B for the remaining main text NCs. Figure S8(C) shows the orthogonality between transition dipoles for 20 representative NCs studied in the main text. The angle between transition dipoles is nearly 90° across all NCs with only 2 fine structure peaks visible, while the observation breaks down naturally for the NCs with 3 fine structure peaks visible. This lends credence to the idea that three peaks are visible when the NC is tilted away from the collection axis, and vice versa when only two peaks are visible.

Polarization-resolved second order correlation spectroscopy

*Measurement uncertainty*

The degree of correlation (DOC) is defined as the ratio between the difference and the sum of co- and cross-polarized correlations,

$$c_\mu = \frac{g^2_{\mu\mu}(0) - g^2_{\mu\mu'}(0)}{g^2_{\mu\mu}(0) + g^2_{\mu\mu'}(0)} \qquad (S1.5)$$

Where $\mu$ and $\mu'$ are orthogonal angles. Each correlation is the normalized center to first side peak ratio

$$g^2(0) = \frac{A_{center}}{A_{side}} = \frac{C}{S} \qquad (S1.6)$$

Both C and S are integrated photon coincidence counts that are Poisson distributed. To compute the degree of correlation, the measured values are directly used (fig. S9). To compute the error on the ratio, the discrete nature of counting statistics means there is no simple closed form on the ratio. We instead sample the underlying distribution.

The underlying coincidence rate is the quantity of interest: given observed k counts and a flat prior on the rate, the Poisson likelihood yields a Gamma-distributed posterior,

$$\lambda|k \sim Gamma(k+1,1) \quad (S1.7)$$

We propagate the uncertainty through the DOC by Monte Carlo: drawing the rate of the center, $\lambda_c$~Gamma$(C+1,1)$ and the side, $\lambda_s$~Gamma$(S+1,1)$, forming the DOC with the underlying distributions, and taking the standard deviation of the resulting distribution as the error, using the mode as the point estimate.(*55*)

*Building the model*

The model is an extension of the published work by Das and Agarwal.(*16*) In their work, Das and Agarwal build an analytical model for polarization correlation functions based on an underlying physical model of a four-level system: one biexciton state, two intermediate paths, and the ground state. The biexciton decays by emission of either a horizontally polarized photon or a vertically polarized photon, in the linear basis, and the remaining excitonic states decay to the ground.

We extend the model to include a third intermediate path whose contribution is either solely as a phonon sink, in the case where the state is invisible, or as a third population with mixed polarization in the case where the state transition dipole tilts into view. We represent this by a rigid-body tilt from normal with polar angles $\Psi, \phi_{rel}$ .

$$Proj(a) = \cos\theta - (1-\cos(\Psi)) cos\phi_{rel} \cos(\theta - \phi_{rel}) \quad (S1.8)$$

$$Proj(b) = \cos\theta - (1-\cos(\Psi)) sin\phi_{rel} \cos(\theta - \phi_{rel}) \quad (S1.9)$$

$$Proj(Z) = \sin\Psi \ \cos(\theta - \phi_{rel}) \quad (S1.10)$$

The core of the model remains identical to that of Das and Agarwal.(*16*) We build in spontaneous emission and incoherent effects via a master equation technique, representing the population contributions by the matrix W and the coherence contributions by the matrix L. We use normalized units for the entire fit to avoid specifying a radiative rate.

$$W = \begin{pmatrix} -(1+\gamma_{ab}+\gamma_{az}) & \gamma_{ba} & \gamma_{za} \\ \gamma_{ab} & -(1+\gamma_{ba}+\gamma_{bz}) & \gamma_{zb} \\ \gamma_{az} & \gamma_{bz} & -(1+\gamma_{za}+\gamma_{zb}) \end{pmatrix} \quad (S1.11)$$

The population matrix acts on each individual population.

$$L = \begin{pmatrix} L_{ab} \\ L_{az} \\ L_{bz} \end{pmatrix} = \begin{pmatrix} 1+\frac{\gamma_{ab}+\gamma_{az}+\gamma_{ba}+\gamma_{bz}}{2} \\ 1+\frac{\gamma_{ab}+\gamma_{az}+\gamma_{za}+\gamma_{zb}}{2} \\ 1+\frac{\gamma_{ba}+\gamma_{bz}+\gamma_{za}+\gamma_{zb}}{2} \end{pmatrix} \quad (S1.12)$$

The coherence matrix acts on every pair of states, i.e. coherences.

While Das and Agarwal retain complete generality of the incoherent terms (Eq 7 in ref (*16*)), we assume incoherent population transfer (EFSR) *via* phonon-assisted spin flip that obeys detailed balance.

$$\gamma_{abs} = rN_B(\Delta, \mathrm{T}) \tag{S1.13}$$

$$\gamma_{em} = r(N_B(\Delta, \mathrm{T}) + 1) \tag{S1.14}$$

We assume a constant coupling term for the reasons set out below.
Our matrix W governs the dynamics of the intermediate state populations. Instead of a closed form solution, we take the numerical approach by matrix inverse (Green's function)

$$\int_0^\infty e^{Wt} dt = -W^{-1} \equiv M \tag{S1.15}$$

So, $M_{ik}$ ends up being the total integrated population that ends up in state 'i' per unit population starting in state k over the course of the entire population decay.
The coherence decay rates are identical and we use the analytical form of the integrated coherence evolution over the population lifetime,

$$G = \int_0^\infty e^{-Lt} \cos(\Delta t - \phi)\, dt \tag{S1.16}$$

Which reduces to an analytical form- for example, the $L_{ab}$ time evolution becomes

$$G_{ab} = \frac{2Proj(A)Proj(B)(L_{ab}cos\phi - \Delta_{ab}sin\phi)}{L_{ab}^2 + \Delta_{ab}^2} \tag{S1.17}$$

We set $\phi = 0$ due to measurement in the linear basis and linear state emission. $\Delta_{ab}$ is simply the energy splitting normalized by $\hbar\Gamma$, where Γ is the exciton radiative rate (identical for both paths). The overall time-averaged second order correlation amplitude for a given state is then given by the projection-weighted sum of the population contributions, given by M, and the coherence contributions, G, over the lifetime decay.
Because the experiment integrates coincidences over the full delay window rather than resolving the correlation in time, the relevant theoretical quantity is the time-integrated correlation $\mathcal{I}(\theta_1, \theta_2) = \int_0^\infty \langle I_{\theta_1}(0) I_{\theta_2}(\tau) \rangle\, d\tau$. This integration is performed analytically: the population dynamics integrate to the Green's matrix $M = -W^{-1}$, and each coherence integrates to the Lorentzian $G_{ij} = L_{ij}/(L_{ij}^2 + \Delta_{ij}^2)$. The degree of correlation is then formed from the co- and cross-polarized integrated correlations as

$$c_\mu = \frac{\mathcal{I}_{\text{co}} - \mathcal{I}_{\text{cross}}}{\mathcal{I}_{\text{co}} + \mathcal{I}_{\text{cross}}} \tag{S1.18}$$

*Coupling rate, r*
We assume a single phonon coupling rate $r = \gamma_0/\Gamma$ common to all three state pairs. This is justified on both physical and statistical grounds. Physically, the acoustic modes available to the nanocrystal form a quasi-continuum above an onset value set by the NC size; the inner transitions ($\Delta_2$ and $\Delta_1$) are sub-meV splittings that sample this same quasi-continuum over a narrow energy window, so their intrinsic coupling strengths are effectively equal. The asymmetry between the pair rates therefore arises not from differing coupling constants but from the thermal phonon occupations $N_B(\Delta, \mathrm{T})$, which we retain per pair.
Statistically, the dataset comprises 8-15 points per nanocrystal; introducing independent couplings would overparametrize the fit, adding degrees of freedom that the data cannot constrain. We therefore adopt a single $r$, with all rate differences carried by the underlying bath phonon occupation $N_B(\Delta, \mathrm{T})$.

*Fine structure splitting*

For the three NCs with no directly visible third state, the presence and position of that state are inferred from a combination of observation and established theory.

- Orthorhombic $CsPbBr_3$ nanocrystals possess a bright exciton triplet of three non-degenerate, linearly polarized state transition dipoles, as established in the literature.(*56*)
- Theory predicts that for near-cube-shaped NCs of orthorhombic crystal structure, the$|Z\rangle$ state lies energetically intermediate between the two bright states $|a\rangle$ and $|b\rangle$ (main text, Figure 2). This level order was established experimentally for orthorhombic CsPbI3 NCs of near-cuboidal shape.(*32*)
- A two-state model does not adequately reproduce the shape of the measured DOC: it overpredicts the correlation of the lower (V) bright state.

Together these motivate the hypothesis that the intermediate energy $|Z\rangle$ state, acting as a phonon-mediated relaxation channel, accounts for the observed DOC shape. The direction of this effect is fixed by the level structure. Because $|Z\rangle$ lies between the other two bright states and the sample is held at 4 K, relaxation into $|Z\rangle$ is governed by the thermal phonon occupation $N_B(\Delta)$ of the connecting transition: the smaller the energy gap between $|Z\rangle$ and one of the other two bright states, the larger $N_B$ and the more strongly $|Z\rangle$ drains that state. Details of the fine structure model are given in Supplementary Text 2 below.

*Fitting the degree of correlation model to the data*

For each nanocrystal we fit the measured degree of correlation $c_\mu(\theta)$to the three-state model by weighted nonlinear least-squares, minimizing $\chi^2 = \sum_i [( c_{\mu,i}^{\mathrm{meas}} - c_\mu(\theta_i))/\sigma_i]^2$, where the $\sigma_i$are the per-point uncertainties from the bootstrap error analysis. The detection amplitudes follow the rigid-body dipole projection of Eqs S1.8-10; three nanocrystals considered in the main text are oriented with their symmetry axis along the optical axis ($\psi = 0$), so the $|Z\rangle$ state transition dipole projects to zero ($A_Z = 0$) and $|Z\rangle$ enters only as a non-radiative relaxation channel, with the detected correlation arising entirely from the $|a\rangle$/$|b\rangle$ bright pair. For NC4 in the main text, the $|Z\rangle$ state becomes visible and contributes to coherence terms to the extent given by the tilt angles, $\Psi$ and $\phi_{rel}$.

Following Das & Agarwal,(*16*) all rates are normalized to the exciton radiative rate $\Gamma$, so the coupling $r = \gamma_0/\Gamma$ is a dimensionless ratio. The radiative rate is not assumed constant across nanocrystals: for each dot, $\Gamma$ is taken from its measured radiative lifetime and enters the model only through the dimensionless splitting $\Delta_{\mathrm{i}}/\Gamma$ in the coherence term.

For the two-peak NCs, the free parameters are the phonon coupling $r$, the transition dipole orientation $\theta_0$, and the $|Z\rangle$-state position $\Delta_2$. A single coupling $r$ is used for all three state pairs, with the pair-to-pair rate differences carried entirely by the thermal occupations $N_B(\Delta, \mathrm{T})$. For the three-peak NC, the free parameters are still the phonon coupling r and the transition dipole orientation $\theta_0$, but now the tilt angles $\Psi$ and $\phi_{rel}$ determine the depolarization as all three fine structure peaks are resolved in energy.

Each fit is initialized by a global search (differential evolution) and refined by a local trust-region least-squares step, which also yields the parameter covariance; this two-stage procedure ensures the reported parameters correspond to the global optimum rather than a local minimum. Parameter uncertainties are obtained from the covariance scaled by $\chi_v^2$, and the shaded confidence bands in the fits are the $\pm 1\sigma$envelopes propagated from this covariance by Monte Carlo sampling.

*Extracted Parameters*
Across the four nanocrystals the fits return a consistent coupling $r \approx 0.6$–$0.7$ and a $|Z\rangle$-state position $\Delta_2 \approx 0.15$–$0.25$ meV, with $\chi_v^2 \approx 2.7$; see Table S1 for all extracted values.

<u>Confined acoustic phonons and exciton fine structure relaxation</u>
*The acoustic phonon spectrum*
The acoustic vibrational modes of the nanocrystal were computed in the continuum-elasticity limit using the Rayleigh-Ritz variational method with the Visscher *xyz* polynomial basis,(39) extended to rounded morphologies through the superquadric formulation of Saviot.(*40*) The displacement field is expanded in monomials $x^l y^m z^n$with $l + m + n \leq N$; we use $N = 10$. The nanocrystal is treated as a free-standing orthorhombic elastic body with edge length $L = 17.8$nm, density $\rho = 4550$kg $m^{-3}$, and the anisotropic stiffness tensor of $CsPbBr_3$, (*38*) whose soft and stiff $z$-shear constants ($C_{44}, C_{55}$) set the energies of the modes relevant to fine-structure relaxation. Mode energies scale as $1/L$, and we report the spectrum across the measured size range $L = 16$–$20$ nm. Corner rounding is parametrized by the superquadric exponent $n$(sharp cube $n \to \infty$).(*40*)
To characterize the modes, we compute the fraction of each mode's total strain energy that is dilatational (volume-changing). This quantity, $f_{comp}$, tells us the compression character of a mode. We find that the lowest modes are all pure shear, $f_{comp} < 0.05$- we plot all shear modes up to the convergence limit in fig. S10. We sketch the region above 350 μeV as a quasi-continuum, since the mode density rises steeply there and individual modes are no longer separately resolved. The shear character of the lowest modes is consistent with our model. Only Z-shear modes have the correct symmetry to couple the in-plane $|a\rangle$ and $|b\rangle$ states to the $|Z\rangle$ state; dilatational modes do not.
Further, the confinement creates a cutoff bounded by the size of the NC. The fundamental mode lies at 115 $\mu eV$ for a sharp cube, below which there are no modes. Notably, the experimental fine structure gaps are close to this limit naturally: NC1 and 2 have inner gaps $\Delta_2$ of just 150 $\mu eV$. Above 350 $\mu eV$, the acoustic modes become a quasi-continuum. The shape of boundary of the NC matters as well- rounded corners stiffen the modes. In fig. S10, the right panel shows the same modes evaluated for slightly rounded NCs. Changing the superquadric exponent to $n = 6.9$ in the Saviot truncation raises all mode energies by $\approx 10\%$. This could provide a mechanism of tuning the overlap between fine structure splitting and available acoustic modes.

*Coupling of acoustic phonons to optical transitions*
By the symmetry of the bright-state transition dipoles, the mode connecting $|a\rangle, |\, b\rangle$ to the out of plane $|Z\rangle$ state is a $z$-shear strain ($\varepsilon_{az}, \varepsilon_{bz}$). This restricts the relevant modes to the $z$-shear set in fig. S10 and removes the compression deformation-potential channel, the large one in most semiconductors. For a confined acoustic mode of wavevector $q$, the squared coupling matrix elements take the standard forms

$$| M_{\mathrm{DP}} |^2 = \frac{\hbar D^2 q}{2V\rho v} \tag{S1.19}$$

$$| M_{\mathrm{PE}} |^2 = \frac{\hbar \, \xi}{V\rho v} \, \frac{A(\hat{q})}{q} \tag{S1.20}$$

where $D$is the (shear) deformation potential, $\xi \propto ( e_{\mathrm{pz}}/\varepsilon_0\varepsilon_r)^2$ the piezoelectric coupling, and $A(\hat{q})$ an angular factor that varies for longitudinal and transverse modes.(*57*) Because the lowest confined modes have the smallest wavevectors ($q \sim \pi/L$), the piezoelectric interaction is favored most strongly for exactly the modes that lie closest to the fine structure splitting. The deformation-potential channel is further suppressed because only the intrinsically small shear deformation potential is symmetry-allowed. We therefore identify piezoelectric coupling as the more plausible mechanism, while noting that the absolute rate is sensitive to parameters not precisely defined for $CsPbBr_3$, chiefly the piezoelectric stress constant $e_{\mathrm{pz}}$ that is not yet determined experimentally, and the dynamic or static dielectric constant.(*58*)

**Supplementary Text 2: Exciton fine structure model**

$CsPbBr_3$ nanocrystals studied in this work possess cuboidal shape with edge lengths in the range 15-17 nm, roughly 5-6 times larger than the bulk exciton radius of $CsPbBr_3$ is $a_x = 3.07\mathrm{nm}$.(*30*) Consequently, neither the assumption of strong quantum confinement (which neglects correlated electron-hole motion), nor the weak confinement approximation (in which the exciton center-of-mass confinement is accounted for but the exciton internal relative electron/hole motion is considered un-modified relative to the exciton in bulk material), provide a good description of the exciton energies and electron-hole exchange overlap.(*59,32)* A faithful description requires an intermediate confinement description, where both the Coulomb correlation of the electron and hole motion and the quantum confinement of the underlying carriers are accounted for.(*59,32)* We now outline a variational approach to this problem. First neglecting electron/hole exchange effects, the effective mass Hamiltonian for the exciton is given by the sum of the electron and hole kinetic energies and their Coulomb interaction energy:

$$\hat{H}_{eff} = -\frac{\hbar^2}{2m_e}\nabla_e^2 - \frac{\hbar^2}{2m_h}\nabla_h^2 - \frac{e^2}{\epsilon_{eff}|\boldsymbol{r}_e - \boldsymbol{r}_h|} \quad . \tag{S2.1}$$

Using this Hamiltonian, we solve the effective mass equation for an exciton within the NC,

$$\hat{H}_{eff} f(\boldsymbol{r}_e, \boldsymbol{r}_h) = E_x \, f(\boldsymbol{r}_e, \boldsymbol{r}_h) \, , \tag{S2.2}$$

subject to the condition that the envelope function vanish on the NC surface. Here, $f(\boldsymbol{r}_e, \boldsymbol{r}_h)$ is the envelope function for the exciton, which is written as a function of the electron and hole coordinates, $\mathbf{r}_e$, $\mathbf{r}_h$, respectively, and $E_x$ is the exciton energy relative to the band gap. The terms $m_e$, $m_h$ are the electron and hole effective masses, respectively, while $\epsilon_{eff}$ is the effective dielectric constant screening the electron/hole Coulomb interaction.(*59*) We apply a variational approach to solving this problem within the parabolic band approximation. The wavefunction of the lowest energy "S" exciton is assumed to be of the form,

$$f(\boldsymbol{r}_e, \boldsymbol{r}_h) = \frac{1}{\sqrt{N(\beta)}} e^{-\beta|r_e - r_h|} \varphi(\boldsymbol{r}_e)\varphi(\boldsymbol{r}_h), \tag{S2.3}$$

where $\beta$ is a parameter reflecting correlation of the electron and hole motion, $\sqrt{N(\beta)}$ is a normalization factor, and $\varphi$ represents lowest energy electron and hole quantum confined levels in a nano-cuboid of edge lengths $L_{x,}L_{y,}L_z$, aligned to the perovskite pseudo-cubic lattice directions, given by,

$$\varphi(x,y,z) = \sqrt{\frac{8}{L_x L_y L_z}}\cos\left(\frac{\pi x}{L_x}\right)\cos\left(\frac{\pi y}{L_y}\right)\cos\left(\frac{\pi z}{L_z}\right). \tag{S2.4}$$

The optimum value of the variational parameter $\beta$ in Eq. S2.3, which we denote $\beta_{opt}$, is determined by minimizing the expectation value of $\widehat{H}_{eff}$, the kinetic plus Coulomb energy, Eq S2.2, using the envelope function Eq. S2.3. The optimization depends on the dimensions of the NC, so that $\beta_{opt}$ is a function of the ledge lengths, i.e., $\beta_{opt} = \beta_{opt}(L_x, L_y, L_z)$. Importantly, the normalization function $N(\beta)$ also becomes a function of the NC edge lengths: $N(\beta_{opt}) = N(\beta_{opt}(L_x, L_y, L_z))$ Once the optimization procedure has been performed, the exchange overlap factor is given by,

$$\Theta = \Theta(L_x, L_y, L_z) = \Omega \int_V d^3\boldsymbol{r}\ |f(\boldsymbol{r},\boldsymbol{r})|^2 = \frac{1}{N(\beta_{opt}(L_x, L_y, L_z))}\frac{27}{8}\frac{\Omega}{L_x L_y L_z}. \tag{S2.5}$$

Plots of this function can be viewed in Suppl. Fig. 22 of Ref. (*32*).

Electron-hole exchange interaction

With a description of the exciton envelope function and energy in the absence of electron-hole exchange we next consider the e/h exchange within perturbation theory. Within an effective mass description, the electron-hole exchange has two parts, the short-range (SR) and long-range (LR) components.

***SR exchange:***

The SR part of the electron-hole exchange interaction can be written in the form of a spin-dependent contact interaction,(*60*)

$$H_{exch}^{SR} = \frac{1}{2}C_{ex}\Omega\left[\hat{I} - (\sigma_e \cdot \sigma_h)\right]\delta(r_e - r_h)\,. \tag{S2.6}$$

In this expression, $C_{ex}$ is the SR exchange constant for the material, $\Omega$ is the volume of the crystal unit cell, $\hat{I}$ is the 4x4 unit matrix, and $\boldsymbol{\sigma_e}, \boldsymbol{\sigma_h}$ are Pauli operators representing the electron and hole spin. To obtain the size-dependence of the SR exchange energies we calculate the exchange interaction over the total exciton wavefunction, including the underlying periodic part, which is given by the product of the band-edge Bloch functions of the conduction and valence bands labelled $i, j$ respectively:

$$\psi_{i,j}(\boldsymbol{r}_e,\boldsymbol{r}_h) = u_i^h(\boldsymbol{r_e})u_j^c(\boldsymbol{r}_h)f(\boldsymbol{r}_e,\boldsymbol{r}_h)\ . \tag{S2.7}$$

Here again, $f(\mathbf{r_e},\mathbf{r_h})$ is the envelope function for the exciton in the intermediate confinement regime given by Eq. S2.3. Integrating over the envelope functions we find that the short-range exchange interaction can be written as an effective spin operator in the form,(*59*)

$$H_{SR} = \frac{1}{2}C_{ex}\,\Theta\left[\hat{I} - \boldsymbol{\sigma}_e.\boldsymbol{\sigma}_h\right]. \tag{S2.8}$$

The term $\Theta$ is the electron-hole exchange overlap factor, representing the probability that the electron and hole reside in the same unit cell:

$$\Theta = \Theta(L_x, L_y, L_z) = \Omega \int_V d^3\boldsymbol{r}\ |f(\boldsymbol{r},\boldsymbol{r})|^2. \qquad \text{(S2.9)}$$

The integration above is taken over the volume, $V$, of the nanocrystal. It is most convenient to express the NC size in terms of an effective length $L_e$ such that $V = L_e^3$. For the ground exciton state in the bulk, the overlap factor is,(*59*)

$$\Theta_{bulk} = \frac{\Omega}{\pi a_x^3}\ , \qquad \text{(S2.10)}$$

where $a_x$ is the bulk exciton Bohr radius. With this result we can rewrite Eq. S2.8 for the ground exciton in terms of the average singlet triplet splitting in the bulk, $\hbar\omega_{st}$,

$$H_{SR} = \frac{3}{4}\ \hbar\omega_{st}\left[\hat{I}\ - \boldsymbol{\sigma}_e.\boldsymbol{\sigma}_h\right]\left(\frac{\Theta(L_x, L_y, L_z)}{\Theta_{bulk}}\right), \qquad \text{(S2.11)}$$

$$\hbar\omega_{st} = 2/3\ C_{ex}\Theta_{bulk}\ . \qquad \text{(S2.12)}$$

This equation is an effective spin operator for the electron-hole exchange interaction, where the size dependence enters in via the electron-hole exchange overlap function, $\Theta$, with is dependent on the NC size through the size-dependence of the exciton envelope function, Eq. S2.3.

To proceed further we need expressions for the electron and hole Bloch functions. The hole Bloch functions $u^h_{J_h,J_{h,z}}(\mathbf{r_e})$have s-like orbital symmetry, which can be represented as (*59,32)*:

$$|\mathrm{u}^h_{1/2,1/2}\rangle = |s\rangle|\uparrow\rangle,$$
$$|u^h_{1/2,1/2}\rangle = |\mathrm{s}\rangle|\downarrow\rangle, \qquad \text{(S2.13)}$$

where the spinor functions ↑ and ↓ are the eigenfunctions of the electron spin projection along z, and the lower-case symbol s denotes an orbital function that transforms as an invariant under the operations of the crystal point symmetry group. The conduction band Bloch functions, $u^c_{J_e,J_{e,z}}(\mathbf{r_h})$,which have p-like orbital symmetry, can be shown to have the form (*59,32)*:

$$|u^e_1\rangle = -\,\mathcal{C}_c\ |c\rangle|\uparrow\rangle - (\mathcal{C}_a\ |a\rangle +\ i\mathcal{C}_b\ |b\rangle)|\downarrow\rangle\ ,$$
$$|u^e_2\rangle = -(\mathcal{C}_a|a\rangle -\ i\ \mathcal{C}_b|b\rangle)|\uparrow\rangle + \mathcal{C}_c|c\rangle|\downarrow\rangle\ . \qquad \text{(S2.14)}$$

In these expressions the lower case symbols $|a\rangle, |b\rangle\ , |c\rangle$ denote p-like orbital functions $p_i$ that transform like *x, y, z* under rotations with the axes *x ,y, z* defined to be aligned to the ***a, b, c*** primitive unit vectors of the orthorhombic lattice. We call attention to the fact that the orthorhombic primitive vectors ***a, b*** are rotated by roughly 45 degrees about the c axis relative to the primitive vectors of the cubic lattice. The factors $\mathcal{C}_a, \mathcal{C}_b, \mathcal{C}_c$ are real numbers. In perovskites of cubic symmetry $\mathcal{C}_a = \mathcal{C}_b, = \mathcal{C}_c = 1/\sqrt{3}$ so that Eq S2.25 are eigenstates of total angular momentum J = ½. In orthorhombic perovskites, $\mathcal{C}_a, \mathcal{C}_b, \mathcal{C}_c$ are generally unequal, reflecting the effect of orthorhombic crystal field splitting. These parameters can be written approximately in terms of two phase angles, $\theta$ and $\phi$, determined by two crystal fields, $\delta$, and $\zeta$. The first, $\delta$, is a "tetragonal" crystal field parameter, which breaks the symmetry between the ***c*** and the ***a, b*** directions; while the second, $\zeta$, is an "orthorhombic" crystal field parameter which reflects the asymmetry between the ***a, b*** directions. The $\mathcal{C}_a, \mathcal{C}_b, \mathcal{C}_c$ parameters are then given as,(*59*)

$$\mathcal{C}_a \approx \mathcal{C}_a(\theta, \phi) = \frac{\cos\phi\cos\theta - \sin\phi}{\sqrt{2}}\ ,$$
$$\mathcal{C}_b \approx \mathcal{C}_b(\theta, \phi) = \frac{\cos\phi\cos\theta + \sin\phi}{\sqrt{2}}\ ,$$
$$\mathcal{C}_c \approx \mathcal{C}_c(\theta, \phi) = \cos\phi\sin\theta\ . \qquad \text{(S2.15)}$$

In these expressions, the phase angle $\theta$ is given in terms of the spin orbit coupling split-off parameter, $\Delta$, and the tetragonal crystal field $\delta$ by (*59*):

$$\tan 2\theta = \frac{2\sqrt{2}\ \Delta}{\Delta - 3\delta}, \qquad \theta \le \frac{\pi}{2}, \tag{S2.16}$$

while the phase angle $\phi$ is determined by,(*59*)

$$\tan 2\phi = \frac{-4\ \zeta\ \cos\theta}{\Delta + \delta + \sqrt{\Delta^2 - \frac{2}{3}\Delta\,\delta + \delta^2}}. \tag{S2.17}$$

We construct an electron-hole pair basis as follows,

$$|P_1\rangle = |u_1^e\rangle|u_1^h\rangle; \quad |P_2\rangle = |u_1^e\rangle|u_2^h\rangle; \quad |P_3\rangle = |u_2^e\rangle|u_1^h\rangle; \quad |P_4\rangle = |u_2^e\rangle|u_2^h\rangle. \tag{S2.18}$$

Using this basis to represent the SR exchange Hamiltonian, Eq. S2.11, and then diagonalizing, we find the exciton fine structure eigenstates $|X_i\rangle$ as $|D\rangle, |A\rangle, |B\rangle, |C\rangle,$ whose transition dipoles to the crystal ground state respectively vanish (*D*), or are aligned along the symmetry directions ***a, b, c*** of the orthorhombic crystal system, with eigenstates denoted by upper case letters *A, B, C* respectively. In this basis the exchange Hamiltonian is represented by,

$$\widetilde{H}_{DABC} = \begin{pmatrix} E_D & 0 & 0 & 0 \\ 0 & E_A & 0 & 0 \\ 0 & 0 & E_B & 0 \\ 0 & 0 & 0 & E_C \end{pmatrix}. \tag{S2.19}$$

In this expression, the exciton eigen energies are given by,(*59*)

$$E_{X_i} = \frac{3}{2}\hbar\omega_{st}\, f_{X_i}\left(\frac{\Theta(L_e)}{\Theta_{bulk}}\right), \tag{S2.20}$$

where for each fine structure level $X_i$ the dimensionless parameter $f_{X_i}$ determines the relative energy order. These are given by,

$$f_D = 0\ , \quad f_A = 2\,C_a^2, \quad f_B = 2\,C_b^2\ , \quad f_C = 2\,C_c^2 \tag{S2.21}$$

It is useful to express the energies and the oscillator strengths in terms of the orthorhombic and tetragonal crystal fields. Following Ref. (*59*), we linearize expressions Eq. S2.15 to Eq. S2.17 in the crystal fields, which is a valid procedure provided that $\delta \ll \Delta$ and $\xi \ll \Delta$. The resulting expressions for the dimensionless functions $f_{X_i}$ are given for each fine structure level in terms of the tetragonal and orthorhombic strains, $\delta, \xi$ by,

$$\begin{aligned} f_D &= 0\ , \\ f_A &= \left(\frac{2}{3} - \frac{4}{9}\frac{\delta}{\Delta_{SO}} + \frac{4}{3}\frac{\xi}{\Delta_{SO}}\right), \\ f_B &= \left(\frac{2}{3} - \frac{4}{9}\frac{\delta}{\Delta_{SO}} - \frac{4}{3}\frac{\xi}{\Delta_{SO}}\right), \\ f_C &= \left(\frac{2}{3} + \frac{8}{9}\frac{\delta}{\Delta_{SO}}\right). \end{aligned} \tag{S2.22}$$

As shown in Ref. (*58*), the relative oscillator strengths for each fine structure level are also proportional to the functions $f_{X_i}$. The transition dipole are given by $\boldsymbol{P}_{X_i} = \langle X_i|\widehat{\boldsymbol{P}}|G\rangle$, evaluated between the exciton state $X_i$, and the crystal ground state $G$, where $\widehat{\boldsymbol{P}}$ is the momentum operator. The corresponding transition dipole matrix elements, are found as $\boldsymbol{P}_{X_i} = P_K\, \mathcal{O}\, \widetilde{\boldsymbol{p}}_{X_i}$, where $P_K$ is the Kane momentum matrix element, $P_K = -\mathrm{i}\,\langle \mathrm{s}|P_c|p_c\rangle$ which in turn is related to the Kane energy $E_p$ by $E_p = 2\,P_K^2/m_0$,(*61*) $\mathcal{O}$ is an overlap integral for the exciton envelope wavefunction, given by, $\mathcal{O} = \int_V f(\boldsymbol{r},\boldsymbol{r})d^3 r$, where the integral is taken over the NC volume, V, and $\widetilde{\boldsymbol{p}}_{X_i}$ are dimensionless vectors describing the orientation of the transition dipole which is determined by the crystal structure, given by,(*32*)

$$\widetilde{\boldsymbol{p}}_A = \sqrt{2}\, \mathcal{C}_a\ \hat{a}\ , \qquad \widetilde{\boldsymbol{p}}_B = \sqrt{2}\, \mathcal{C}_b\ \hat{b}\ , \qquad \widetilde{\boldsymbol{p}}_C = \sqrt{2}\, \mathcal{C}_c\ \hat{c}\ . \tag{S2.23}$$

Here, $\hat{a}, \hat{b}$, and $\hat{c}$ are unit vectors aligned to the primitive orthorhombic lattice vectors, ***a, b, c***.

***Long-range exchange***

In addition to the SR exchange interaction discussed above, the electron-hole exchange interaction contains a "long range" piece which significantly impacts the exciton fine structure as this term is sensitive to the overall shape of the NC and has a magnitude that is significantly larger than the SR exchange piece. The LR exchange corrections were shown by Cho to be equivalent to the self-energy of the exciton polarization.(*62*) According to Cho, a given exciton state, $i$, with a non-vanishing electric dipole transition matrix element is accompanied by an electric dipole transition moment density $\boldsymbol{\wp}_i(\boldsymbol{r})$. In our context, the transition dipole density associated with exciton state $X_i$ is given by,(*60*)

$$\boldsymbol{\wp}_i(r_e) = \frac{e|\,P_K|}{m_0\omega_i}\int d^3\,r_h\, f(r_e,r_h)\,\widetilde{\boldsymbol{p}}_i\,\delta(r_e - r_h) \approx \frac{e\,|P_K|}{m_0\omega_0} f(r_e,r_e)\,\widetilde{\boldsymbol{p}}_i\,. \tag{S2.24}$$

Here, $f(r_e,r_h)$ is the exciton envelope function; the $\widetilde{\boldsymbol{p}}_i$ are the dimensionless electric transition dipole vectors (given in the $D, A, B, C$ basis by Eq. S2.23); $\omega_{X_i}$ is the angular frequency of the exciton resonance. On the right of this equation, we approximate $\hbar\omega_i$ by the exciton energy $\hbar\omega_0$ evaluated without fine structure corrections. Owing to its spatial variation, the transition dipole density has an associated effective charge density $\rho_i(\boldsymbol{r}) = -\nabla\cdot\boldsymbol{\wp}_i(\boldsymbol{r})$. The long-range exchange interaction can be written as the Coulomb energy of this effective charge density (*63*):

$$H_{i,j}^{LR} = \int_{V_1} dV_1 \int_{V_2} dV_2\ \ [-\boldsymbol{\nabla}_1\cdot\boldsymbol{\wp}_i(\boldsymbol{r_1})]^*\, V(\boldsymbol{r_1}-\boldsymbol{r_2})\big[-\boldsymbol{\nabla}_2\cdot\boldsymbol{\wp}_j(\boldsymbol{r_2})\big]. \tag{S2.25}$$

In the integral, $V(\boldsymbol{r_1},\boldsymbol{r_2})$ is the Coulomb interaction potential between the effective charge at positions $\boldsymbol{r_1}$ and $\boldsymbol{r_2}$. Given multiple sublevels $i$, this interaction will, in general, couple the sub-levels.

Since the QD bounding facets belong to the $\{100\}_c$ pseudocubic planes, the $C$ exciton transition dipole is normal to top and bottom NC bounding facets, while the $A$ and $B$ exciton transition dipoles are oriented approximately 45 degrees from the lateral bounding surfaces.(*32*) This is a consequence of the $\sqrt{2}\times\sqrt{2}\times 2$ orthorhombic modification characteristic of inorganic metal halide perovskite. Following the analysis in Ref. *32* and Ref. *63*, the LR exchange Hamiltonian is represented as a 4x4 matrix in the $D, A, B, C$ basis by,

$$\boldsymbol{H}_{LR}{}^{ex} = \frac{3}{2}\left\{\ \frac{\hbar\omega_{LT}}{3}(1+g_\kappa)\,\widetilde{\boldsymbol{A}}\right\}\left(\frac{\Theta(L_x,L_y,L_z)}{\Theta_{bulk}}\right), \tag{S2.26}$$

where the D state correction is zero and there is no coupling between the D state and the bright exciton $X, Y$, and $Z$ states. In Eq. S2.26, $\hbar\omega_{LT}$ is the bulk longitudinal/transverse exciton splitting, $g_\kappa = \frac{12}{\pi^2}\left(\frac{\kappa-1}{\kappa+2}\right)$ is a factor that accounts for enhancement of the LR exchange interaction due to image charge effects, parameterized by $\kappa = \epsilon_\infty^{NC}/\epsilon_\infty^{med}$, the ratio of the high frequency dielectric constant in the QD to that of the surrounding medium, $\Theta_{bulk} = \frac{\Omega}{\pi a_X^3}$ is the exchange overlap factor for bulk excitons. The dimensionless matrix $\widetilde{\boldsymbol{A}}$ reflects the effects of the geometrical anisotropy of the NC shape and the orientations of the electric dipole transition vectors for each state $n$ within the set $X, Y, Z$. Its matrix element are,(*63*)

$$A_{n,m} = \tilde{p}_{n,x}{}^{*}\, \tilde{p}_{m,x}\, \mathcal{A}_{\mathrm{X}} + \tilde{p}_{n,y}{}^{*}\, \tilde{p}_{m,y}\, \mathcal{A}_{\mathrm{Y}} + \tilde{p}_{n,z}{}^{*}\, \tilde{p}_{m,z}\, \mathcal{A}_{\mathrm{Z}}, \tag{S2.27}$$

where $\tilde{p}_{m,i}$ are the components of the dimensionless electric transition dipole vectors from Eq. S3.27 in a coordinate system $x, y, z$ whose axes are aligned to the NC edge directions, $X, Y, Z$ rather than the symmetry axes of the lattice. The functions $\mathcal{A}_{X_i}$ in Eq. S6.8 are dimensionless shape functions. Plots of these functions can be viewed in Suppl. Fig. 26 of Ref. *32*. They are defined as,

$$\mathcal{A}_{X_i} = \mathcal{A}_{X_i}(L_x, L_y, L_z) \equiv \frac{3}{4\pi} \frac{\Omega\, I_{X_i}(L_x, L_y, L_z)}{\Theta(L_x, L_y, L_z)}, \tag{S2.28}$$

which are written in terms of Coulomb integrals $I_{X_i}(L_x, L_y, L_z)$ given in terms of the envelope functions $f(\boldsymbol{r}_e, \boldsymbol{r}_h)$ of Eq. S6.1 by,(*59,32*)

$$I_{X_i} = \iiint_{-\frac{L_x}{2}\,-\frac{L_y}{2}\,-\frac{L_z}{2}}^{\frac{L_x}{2}\,\frac{L_y}{2}\,\frac{L_z}{2}} \mathrm{d}^3\boldsymbol{r}_1 \iiint_{-\frac{L_x}{2}\,-\frac{L_y}{2}\,-\frac{L_z}{2}}^{\frac{L_x}{2}\,\frac{L_y}{2}\,\frac{L_z}{2}} \mathrm{d}^3\boldsymbol{r}_2 \left[\frac{\mathrm{d}f(\boldsymbol{r_1}, \boldsymbol{r_1})}{\mathrm{d}\boldsymbol{r}_{1,i}}\right]^{*} \frac{1}{|\boldsymbol{r_1} - \boldsymbol{r_2}|} \left[\frac{\mathrm{d}f(\boldsymbol{r_2}, \boldsymbol{r_2})}{\mathrm{d}\boldsymbol{r}_{2,\mathrm{i}}}\right]. \tag{S2.29}$$

Note that for a general cuboidal shape, the basal edge lengths $L_x \neq L_y$. Consequently, the fine structure levels with electric dipole transition matrix elements with components in the $x, y$ directions are coupled by the LR exchange and consequently mix.(*32*) The exception to this is for the special case of a square basal plane, $L_x = L_y$.

Putting it all together the exciton fine structure in the case of pseudo-cubic facets is given by the following expressions: The dark *D* exciton, and the bright $\hat{c}$-polarized *C* exciton have energies given by,

$$E_{\mathrm{D}} = 0\,, \tag{S2.30}$$

$$E_{\mathrm{C}} = \frac{3}{2}\left\{\hbar\omega_{ST} + \frac{\hbar\omega_{LT}}{3}\left[1 + \frac{12}{\pi^2}\left(\frac{\kappa - 1}{\kappa + 2}\right)\right]\mathcal{A}_{\mathrm{C}}\right\} f_{\mathrm{C}}\left(\frac{\Theta(L_e)}{\Theta_{bulk}}\right). \tag{S2.31}$$

The remaining two excitons are formed from the coupled *A* and *B* excitons and are determined by diagonalizing the following Hamiltonian:

$$\widetilde{\mathrm{H}}_{\mathrm{A,B}}{}^{\mathrm{ex}} = \left\{\frac{3}{2}\hbar\omega_{ST}\begin{pmatrix}\mathrm{f_A} & 0\\ 0 & \mathrm{f_B}\end{pmatrix} + \frac{\hbar\omega_{LT}}{3}\left[1 + \frac{12}{\pi^2}\left(\frac{\kappa - 1}{\kappa + 2}\right)\right]\begin{pmatrix}\mathrm{f_A}\mathcal{A}_{\mathrm{A,A}} & \sqrt{\mathrm{f_A f_B}}\mathcal{A}_{\mathrm{A,B}}\\ \sqrt{\mathrm{f_A f_B}}\,\mathcal{A}_{\mathrm{BA}} & \mathrm{f_B}\mathcal{A}_{\mathrm{BB}}\end{pmatrix}\right\}\left(\frac{\Theta(L_e)}{\Theta_{bulk}}\right). \tag{S2.32}$$

The coupled eigenstates correspond to the pure *A,B* excitons only in the case that the NC edge lengths $L_x, L_y$ along the pseudo-cubic axes $[100]_c$ and $[010]_c$, respectively, are equal: $L_x = L_y$, since in that configuration, the cross-coupling term $\mathcal{A}_{A,B}\ \mathcal{A}_{B,A}$ vanishes. In all other cases, the exciton states are a superposition of the *A, B* excitons with transition dipoles whose orientation relative to the NC facets changes with the aspect ratio $L_x/L_y$ .

***Parameterization***:

Material parameters necessary to compute the exciton fine structure using Eqs. S2.31-S2.32 are all available in the literature (*30*) with the exception of the tetragonal and orthorhombic strains, $\delta, \xi$ which reflect the orthorhombic distortion in cuboidal NCs with average size L = 16.8 nm (Fig S5C). The crystal fields are determined fitting fine structure splitting data in **Table S1** to the model, Eqs. S2.31-S2.32, assuming that the average shape is that of a perfect cube, based on image analysis of transmission electron micrographs; the squareness distributions shown in fig. S5E, indicate that the mean particle shape is square, not rectangular, in profile. Model parameters are summarized in **Table S2** and used to compute the median exciton fine structure shown in main text, **Figure 2**.

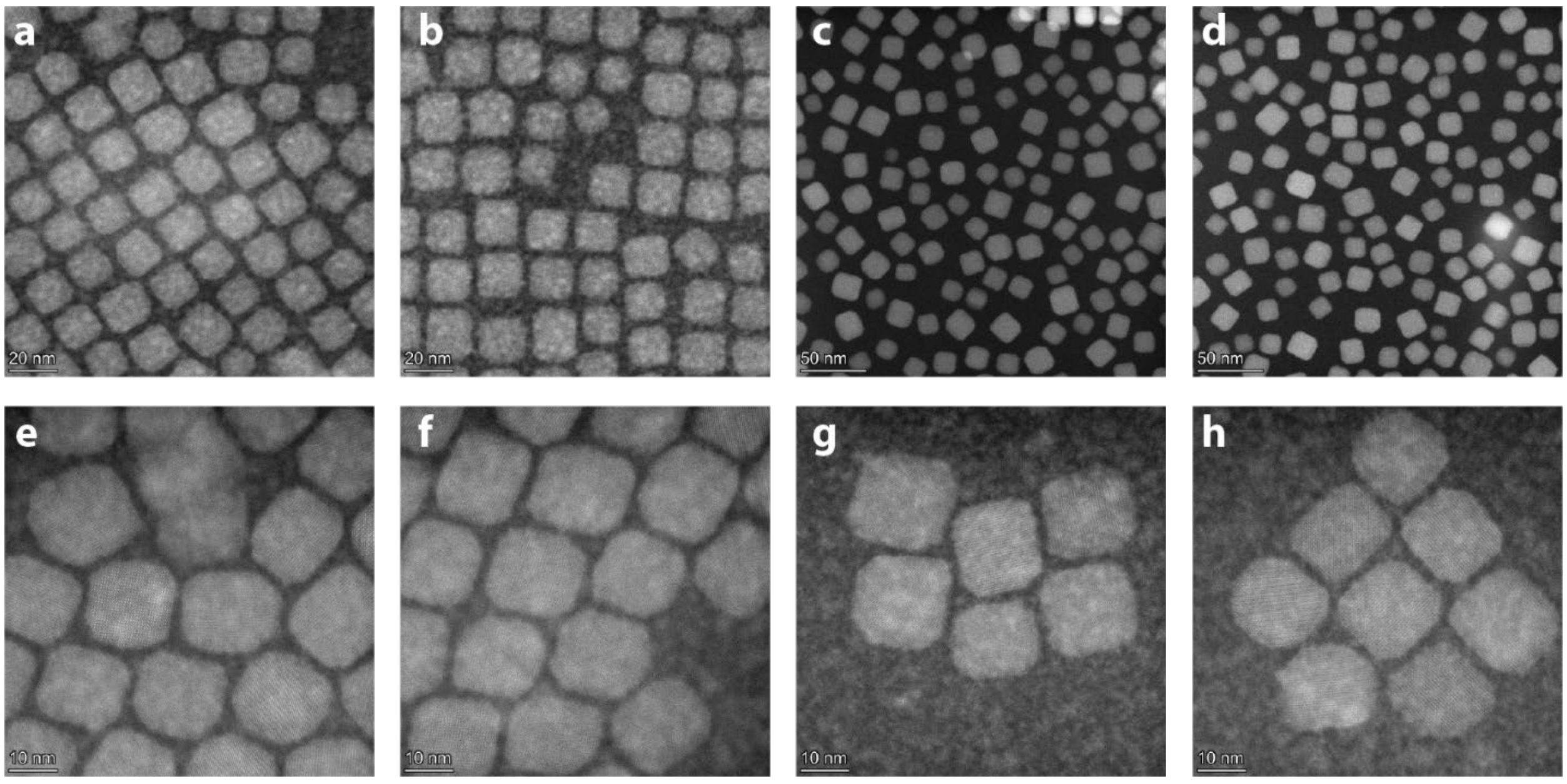

**fig. S1.**

Selected high angle annular dark field scanning transmission electron micrographs (HAADF STEM). Across the sample, the nanoparticles are largely square with rounded edges and an average edge length of 17 nm.

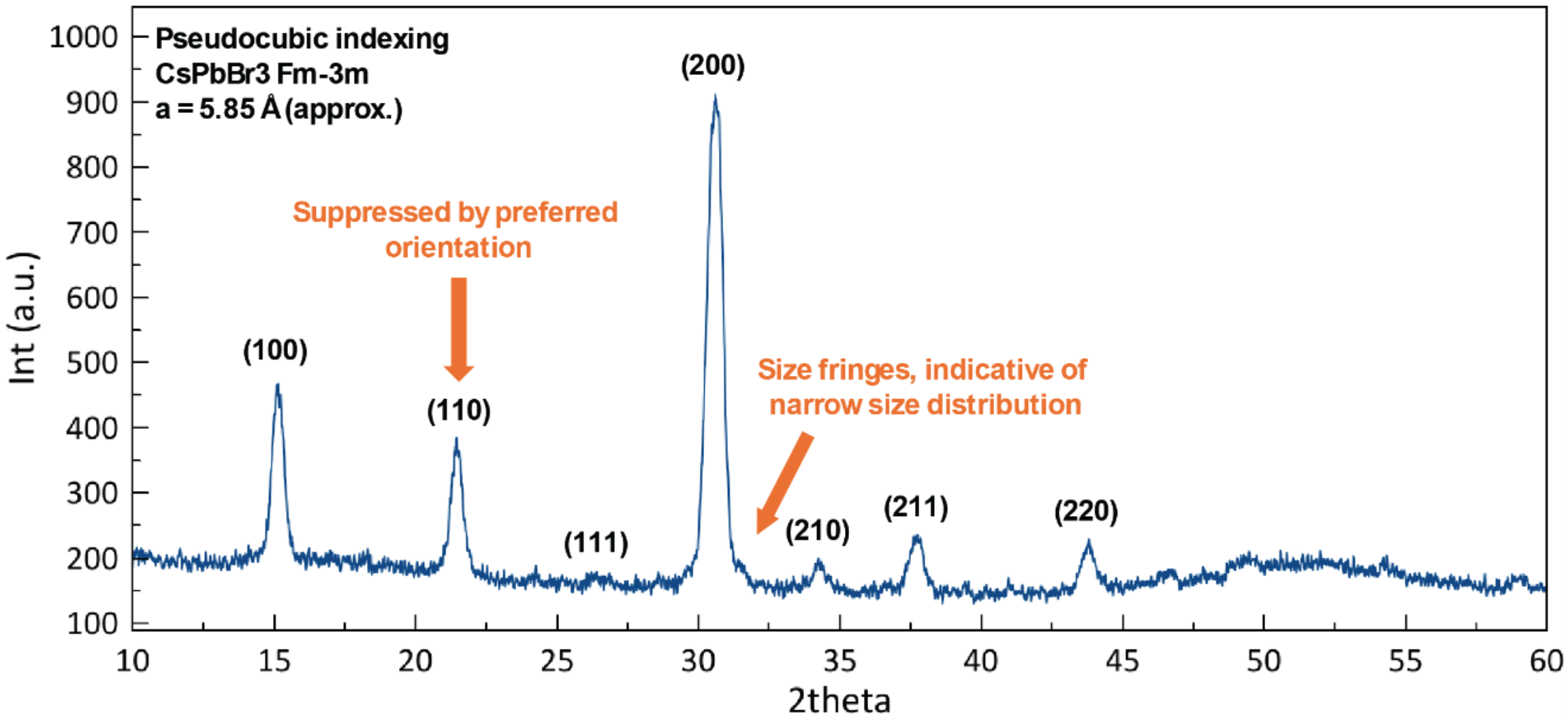

**fig. S2.**

Powder XRD pattern of drop-cast $CsPbBr_3$ nanocrystals. Reflections are indexed using a pseudo-cubic unit cell for simplicity. The enhanced intensities of the (100) and (200) reflections relative to the other peaks indicate preferential orientation, as expected from the cubic morphology of the nanocrystals. In addition, the weak features at the base of the (200) reflection are consistent with first-order finite-size Laue fringes, compatible with a narrow particle-size distribution.(*70*)

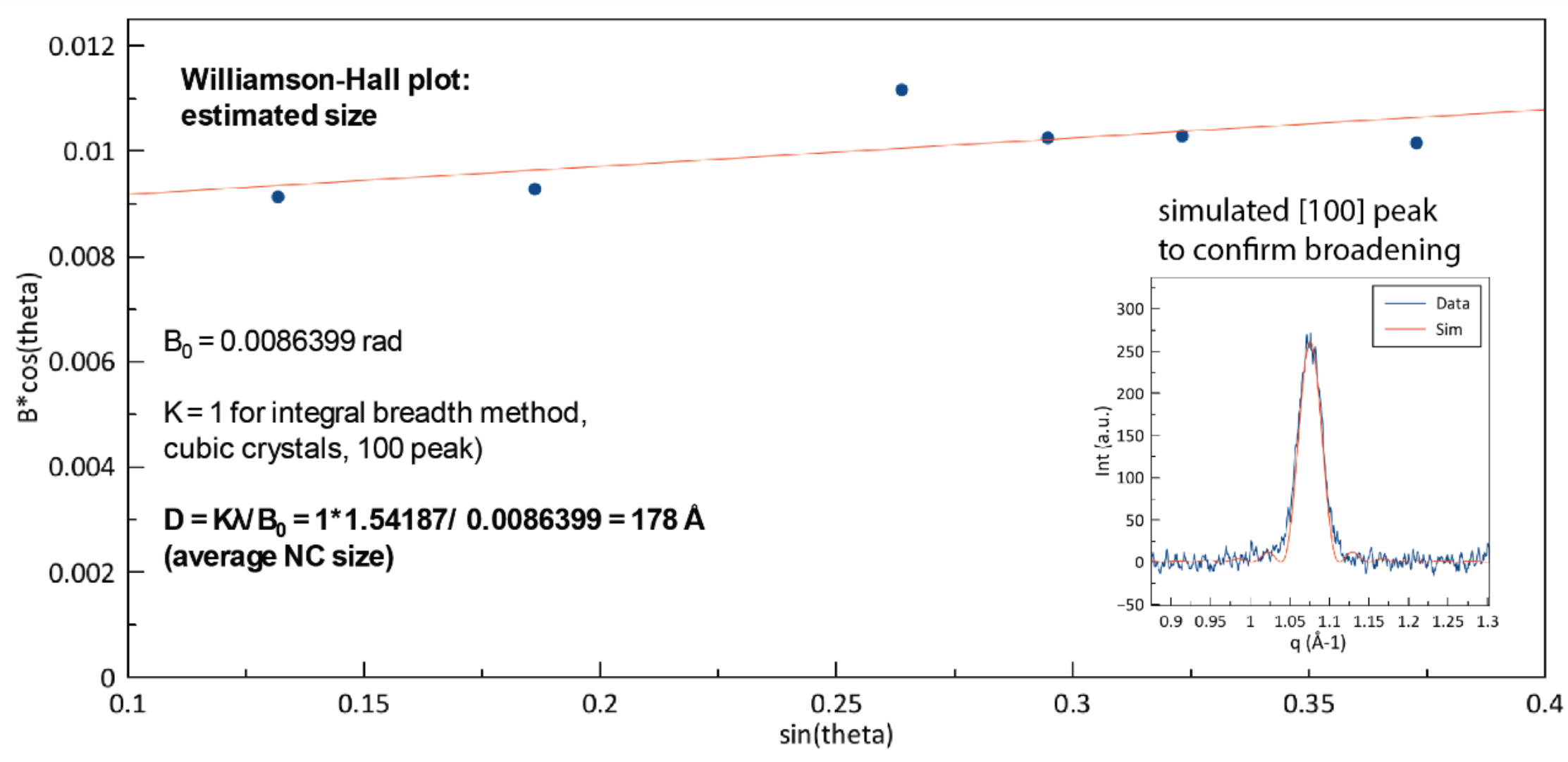

**fig. S3.**

Williamson-Hall analysis of the peak widths in the pattern shown in Figure S2. The analysis yields an average coherent domain size of 17.8 nm, in good agreement with the size estimated by TEM. The inset compares the experimental (100) peak with a peak profile simulated for the corresponding domain size.

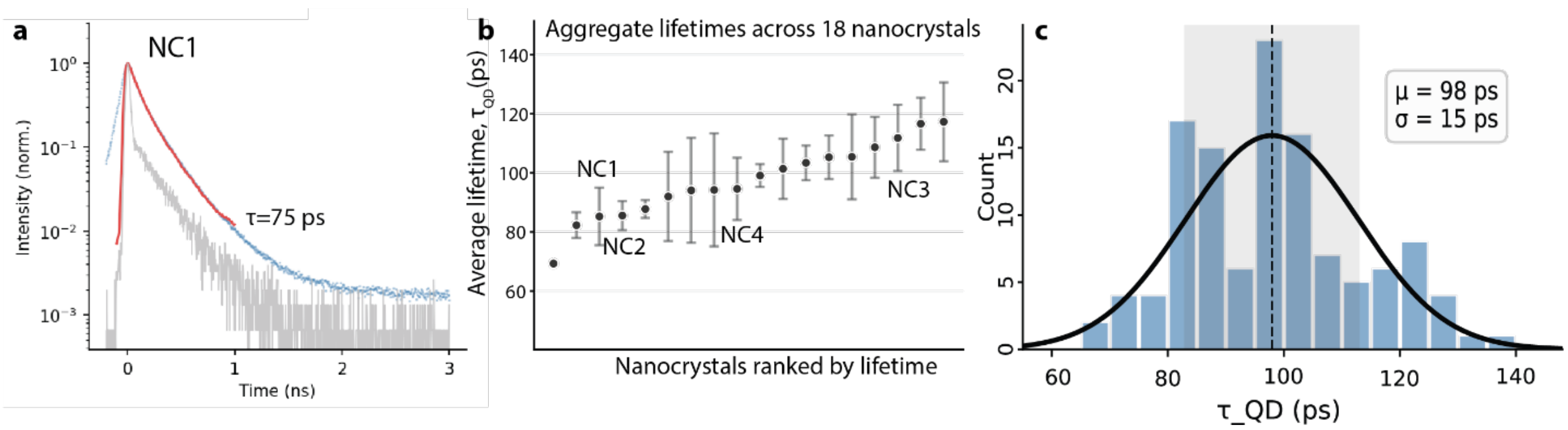

**fig. S4.**

Single-nanocrystal on-state lifetimes. (A) Individual lifetimes were fit *via* reconvolution with the instrument response function (IRF). The fit is shown in red, while the data is shown in blue, and the IRF in gray. (B) Aggregate lifetimes across 18 nanocrystals measured. Error bars are shown where repeat measurements are taken. (C) The distribution of lifetimes across the sample follows a Gaussian distribution centered at 100 ps ± 15 ps.

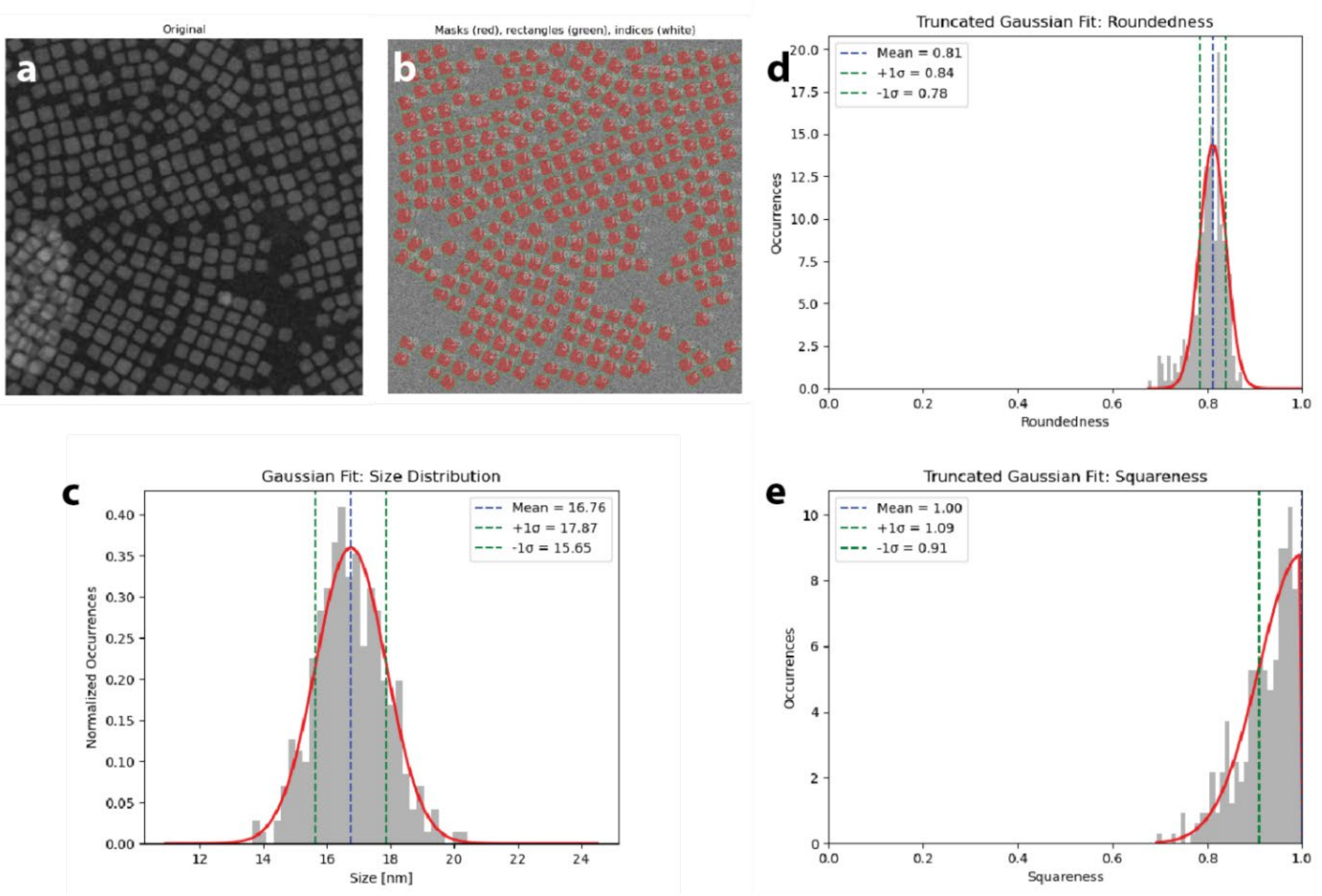

**fig. S5.**

Image analysis of transmission electron micrographs. (A) High angle annular dark field scanning transmission electron micrograph shown in the main text. (B) Image segmentation applied to the original identifies individual particles. (C) The edge length distribution across the particles fits well to a Gaussian with mean size of 16.8 nm $\pm$ 0.1 nm. (D) The particle roundedness, determined by the area deviation from a bounding rectangle. The particles are slightly rounded. (E) A truncated Gaussian fit to show the mean particle shape is square, not rectangular, in profile.

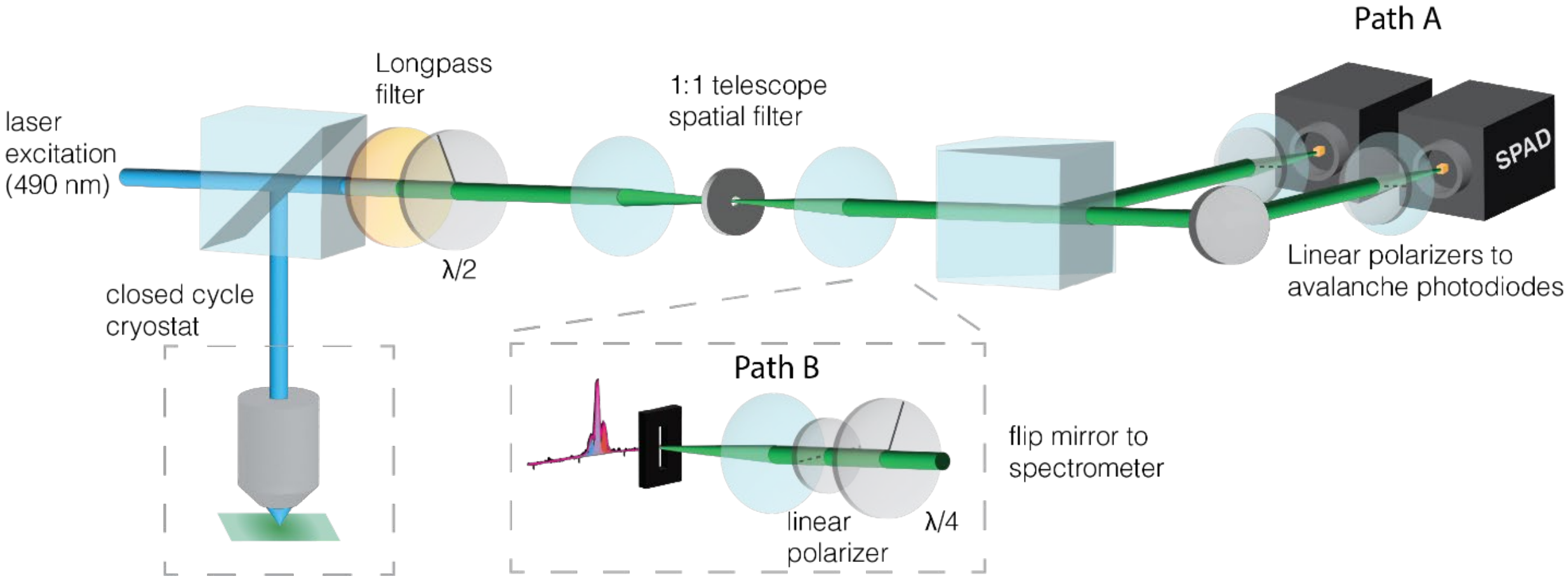

**fig. S6.**
A schematic of the homebuilt confocal microscope used. Key features are the use of polarization maintaining mirrors centered at 532 nm, a 0 degree longpass filter, as well as the use of a half-wave plate to select the nanocrystal emission basis. The setup has two paths, A and B; the former enables the measurement of polarization-resolved second order correlations while the latter, energy-resolved Stokes parameters.

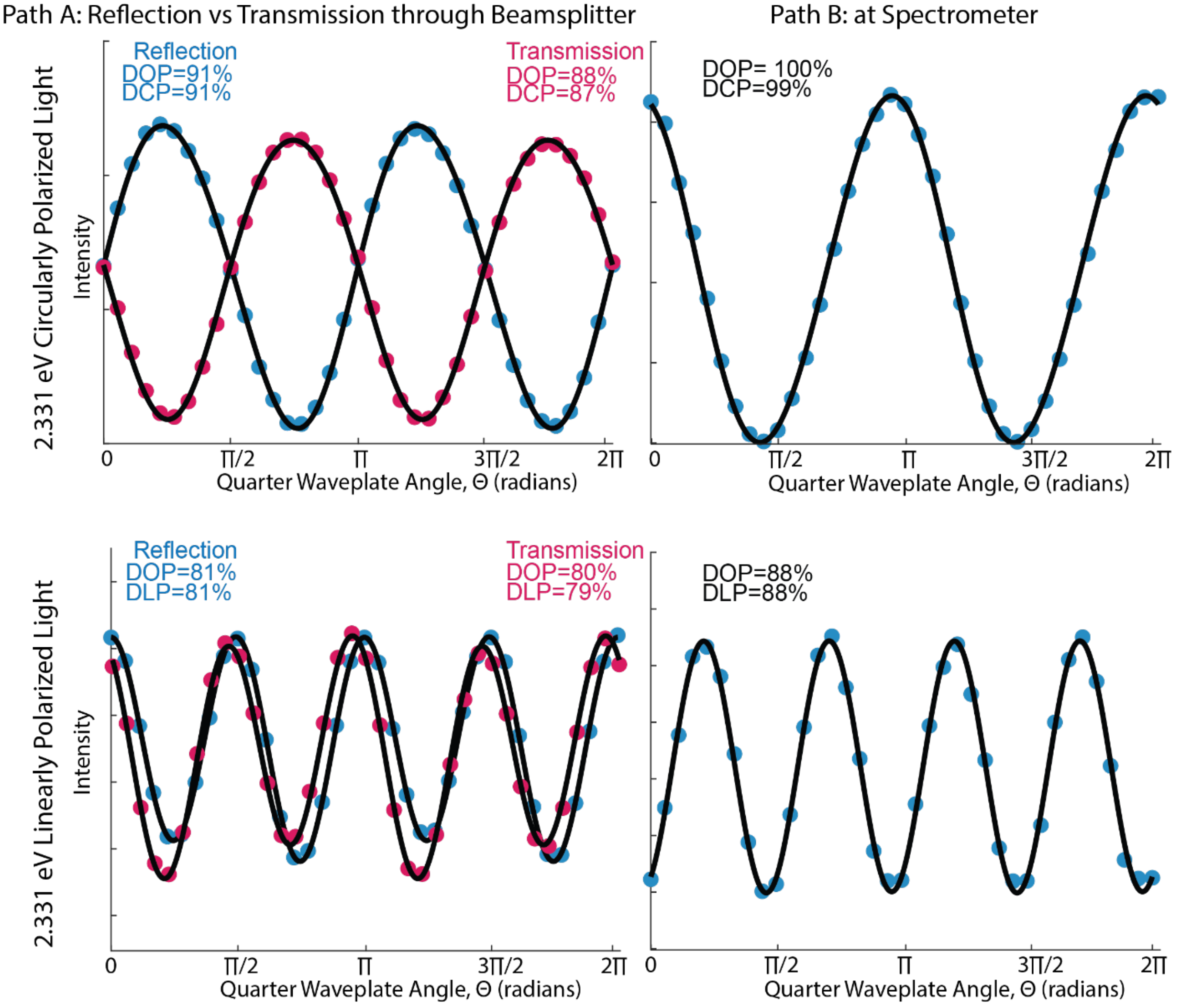

**fig. S7.**

Depolarization of the optical setup. Stokes parameters were measured at the detection at path A and B. The left-hand panels show the measured Stokes parameters at the single photon detectors, both through the transmission and the reflection paths of the final beamsplitter. The right-hand panels show the measured Stokes parameters at the spectrometer. The upper panels show the maintenance of circularly polarized light, while the lower panels show the maintenance of linearly polarized light.

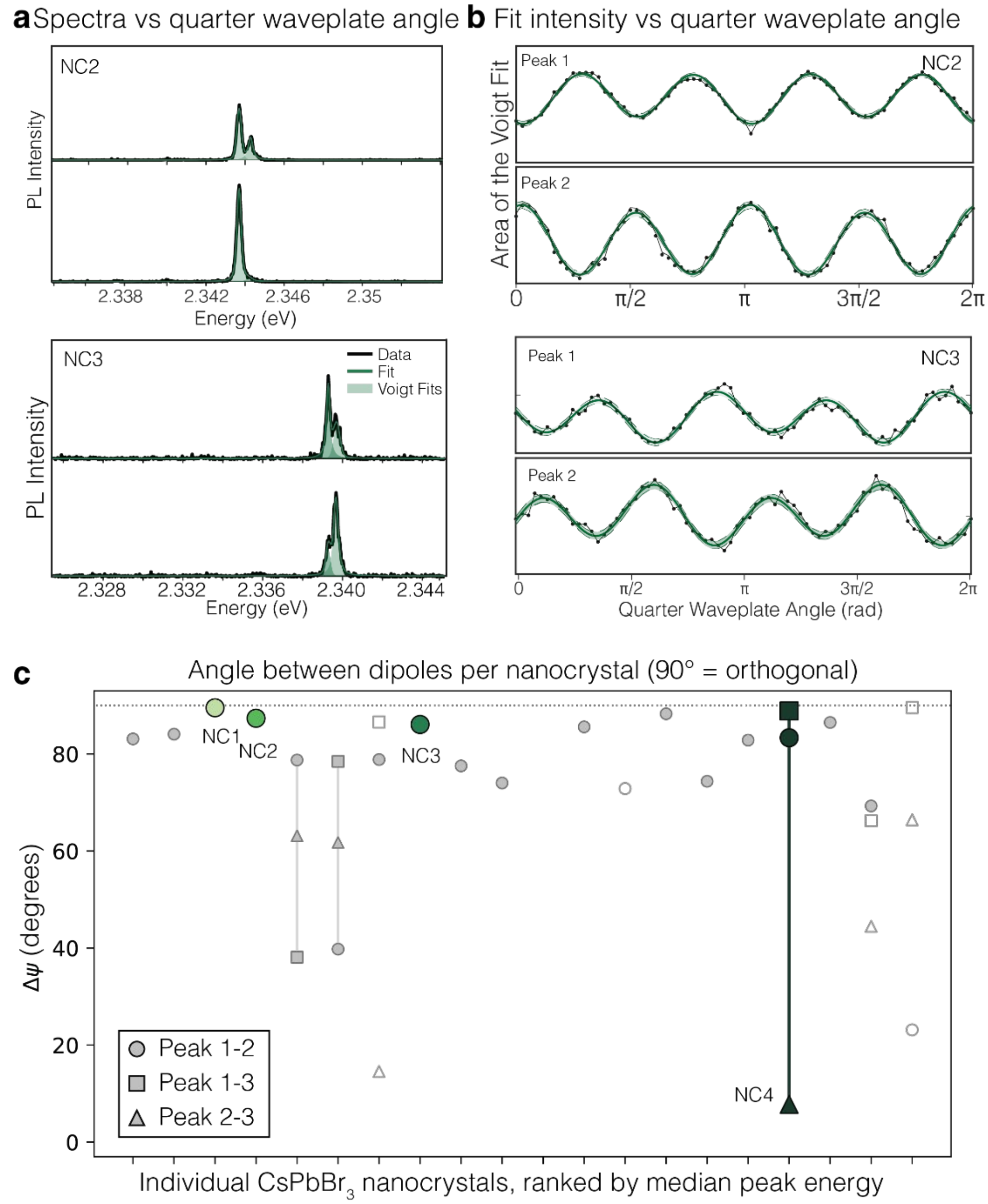

**fig. S8.**

Polarization analysis of the remaining NC2 and NC3 from the main text. (A) shows the as-measured PL spectra for NC 2 and 3 with varying angles of the rotating quarter waveplate analyzer. Each underlying peak is fit to a Voigt profile and the sum fit is overlaid on the data. (B) Integrated areas of the Voigt fits for each resolved peak in Panel (A) as a function of quarter waveplate angle. (C) The complementary angles between transition dipoles across the sample of 20 nanocrystals. As expected, nanocrystals with only 2 exhibited peaks show an angle of 90 degrees, orthogonal. Nanocrystals with 3 exhibited peaks deviate from orthogonality due to the projection of the third state transition dipole into the plane; however, all retain at least one highly orthogonal pair.

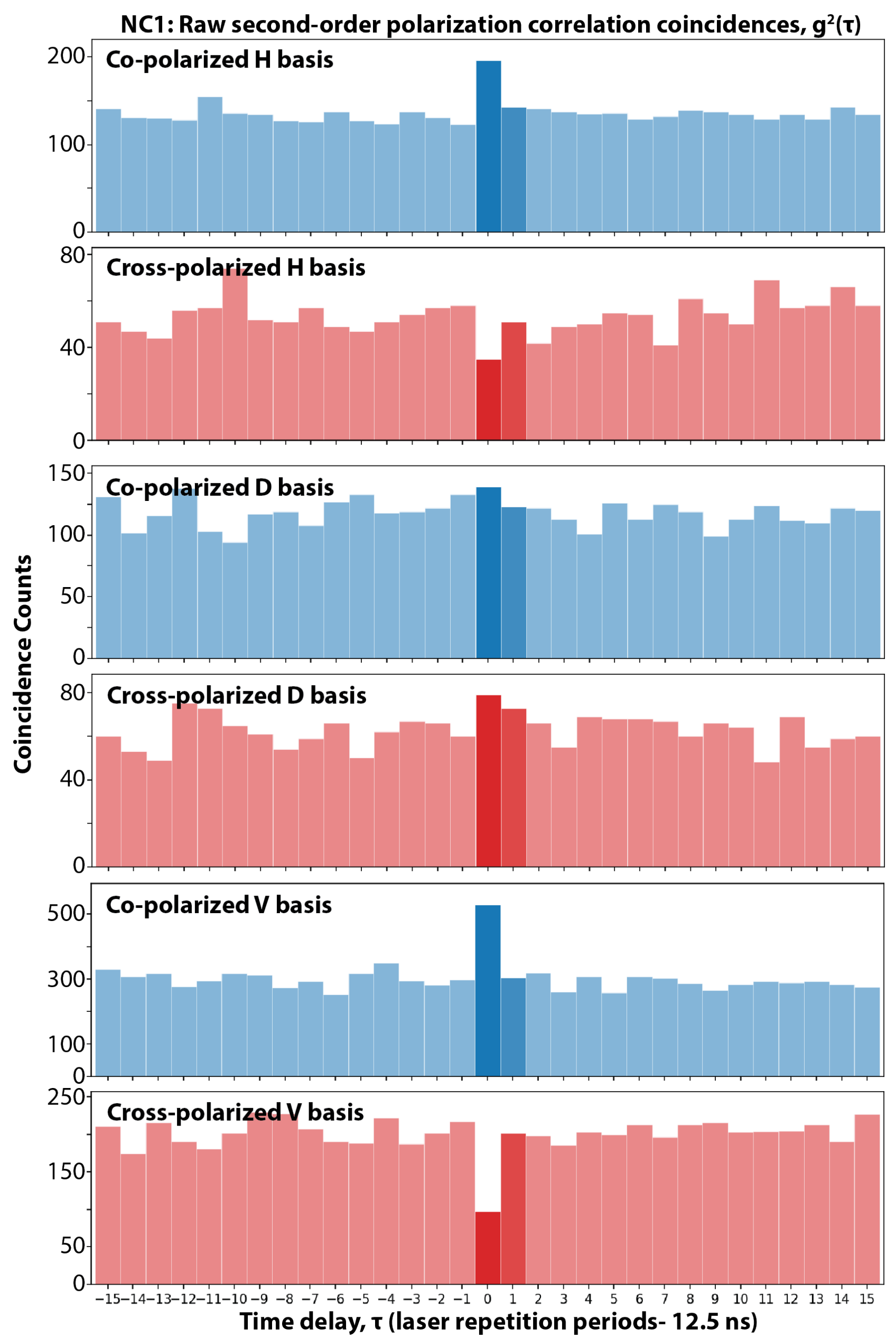

**fig. S9.**

Example raw correlation data for NC1 from the main text. Co- and cross-polarized second order correlations are shown at H, D, and V polarization basis, with 12.5 ns bins corresponding to the laser repetition rate. The center peak corresponds to the zero-time correlation, or the two-photon events occurring within the same laser pulse (a biexciton). This peak is normalized to the average of the nearest 15 side peaks. A clear pattern appears: when the measurement basis aligns with the emitting state transition dipoles, the co-polarized center peak reaches a maximum while the cross-polarized center peak, a minimum. The cross-polarized peak never reaches zero due to exciton fine structure relaxation. Unpolarized background counts or depolarization from the objective would increase counts equally across polarization basis and pulse.

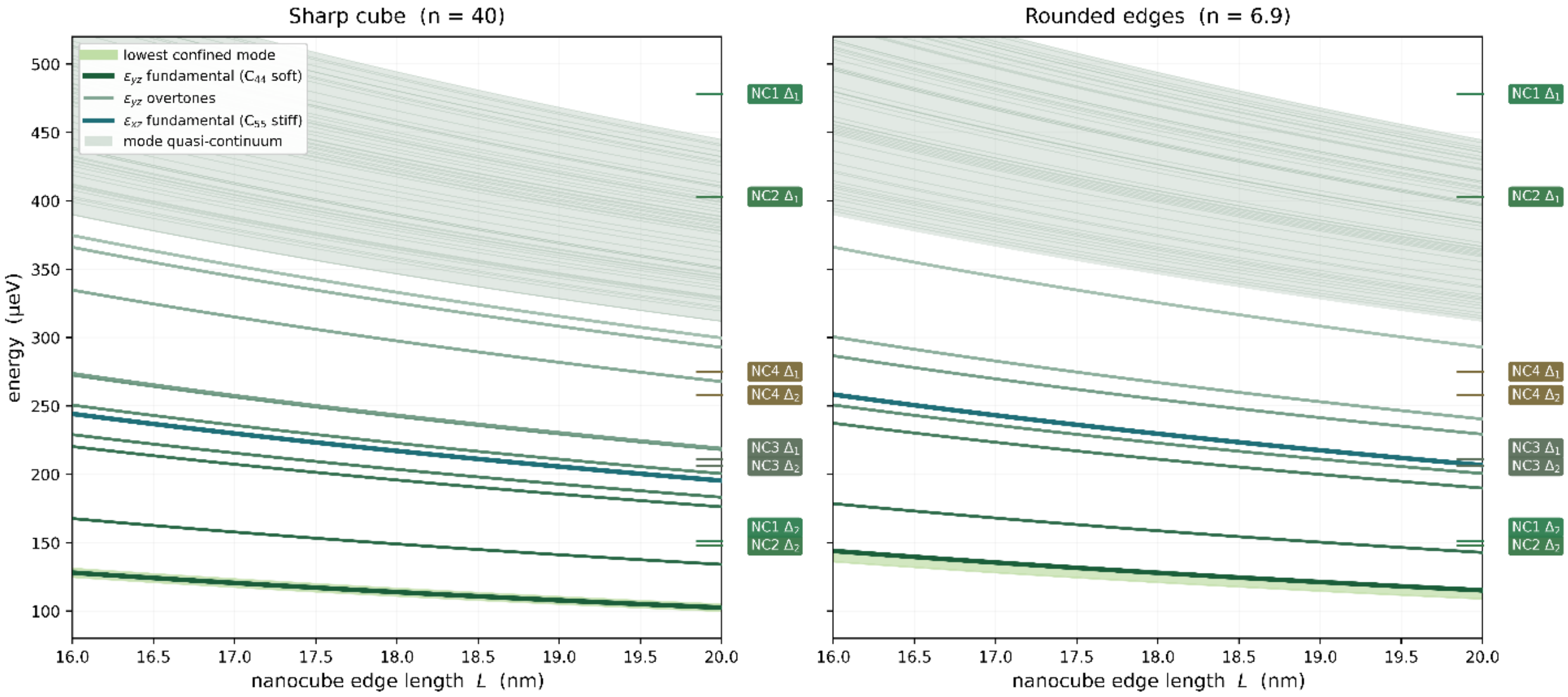

**fig. S10.**

Acoustic phonon mode energies versus particle edge length using the xyz algorithm for free cubes,(*39*) with an underlying orthorhombic *Pnma* symmetry. The fine structure splittings of the four main text nanocrystals are labeled on the side for reference. The lowest-lying mode is bounded by the nanocrystal edge length. Further, all lowest-lying modes are shear for the confined orthorhombic lattice. The left panel shows the calculation for a perfectly sharp cube, according to the method described by Saviot,(*40*) while the right for a rounded cube similar to those measured in this experiment. Rounding the edges increases the stiffness of the acoustic modes by up to 10%.

**Table S1.**
Extracted parameters from the fit of the second order correlations to the three-state phonon relaxation model for NC1-4.

| | Measured lifetime (ps)* | Outer energy splitting, meV ($\Delta_1 + \Delta_2$)* | Lower energy splitting, meV ($\Delta_1$) | Upper energy splitting, meV ($\Delta_2$) | EFSR ratio, $r = \gamma_0/\Gamma$ | Off-axis tilt angles, $\Psi_{tilt}$ / $\phi_{rel}$ | Goodness of fit, reduced $\chi^2$ ($\chi_\nu^2$) | Maximum coherence ratio, $T_2/2T_1$, of lowest state at 4K |
|---|---|---|---|---|---|---|---|---|
| NC1 | $85 \pm 10$ | 0.629 | $0.48 \pm 0.03$ | $0.15 \pm 0.03$ | $0.69 \pm 0.08$ | - | 2.96 | $0.81 \pm 0.03$ |
| NC2 | $85 \pm 5$ | 0.551 | $0.40 \pm 0.03$ | $0.15 \pm 0.03$ | $0.66 \pm 0.09$ | - | 2.66 | $0.77 \pm 0.03$ |
| NC3 | $112 \pm 11$ | 0.417 | $0.21 \pm 0.03$ | $0.21 \pm 0.03$ | $0.62 \pm 0.07$ | - | 2.61 | $0.57 \pm 0.05$ |
| NC4 | $95 \pm 11$ | 0.533 | 0.275* | 0.258 * | $0.60 \pm 0.05$ | 42° / -7° | 1.78 | $0.67 \pm 0.02$ |

* directly measured

**Table S2**. Material parameters necessary to compute the exciton fine structure using Eqs. S2.31-S2.32.

| Parameter | Value | Source |
|---|---|---|
| Exciton reduced effective mass $\mu/m_0$ | 0.126 | Ref. *64* |
| Exciton effective dielectric constant $\epsilon_{eff}$ | 7.3 | Ref. *64* |
| Exciton Bohr radius in bulk $CsPbBr_3$, $a_x$ | 3.07 nm | $a_x = a_o\, \epsilon_{eff}/\mu$ where $a_o$ is the hydrogen Bohr radius |
| Exciton binding energy in bulk $CsPbBr_3$, $B_x$ | 32 meV | $B_x = \frac{\hbar^2}{2\mu a_x^2}$ |
| Bulk ST splitting, $\hbar\omega_{ST}$ | 0.49 meV | Calculated using Eq. S2.12 with parameters from Ref. *65* |
| Bulk LT splitting, $\hbar\omega_{LT}$ | 5.4 meV | Ref. *66* |
| Spin-orbit split-off parameter, $\Delta$ | 1.2 eV | Ref. *67* |
| $CsPbBr_3$ high frequency dielectric constant $\epsilon_{NC,\infty}$ | 4.3 | Ref. *68* |
| PMMA dielectric $\epsilon_{med} = n^2$ at 530 nm | 2.22 | Ref. *69* |
| Dielectric contrast ratio, $\kappa = \epsilon_{NC,\infty}/\epsilon_{med}$ | 1.68 | $\kappa = \epsilon_{NC,\infty}/\epsilon_{med}$ |
| Tetragonal crystal field, $\delta$, in nanocrystals | 17 meV | This work |
| Orthorhombic crystal field, $\xi$, in nanocrystals | 56 meV | This work |